\documentclass[twocolumn,tighten]{aastex62}
\usepackage{natbib}

\usepackage[hang,raggedright,tight]{subfigure}
\usepackage{longtable}
\usepackage[figuresright]{rotating}
\usepackage{float}
\usepackage[toc,page]{appendix}

\usepackage{gensymb}
\usepackage{threeparttable}
\usepackage{diagbox}
\usepackage{pbox}
\usepackage{multirow}  

\def\gsim{\;\rlap{\lower 2.5pt
\hbox{$\sim$}}\raise 1.5pt\hbox{$>$}\;}
\def\lsim{\;\rlap{\lower 2.5pt
\hbox{$\sim$}}\raise 1.5pt\hbox{$<$}\;}

\usepackage{xcolor}
\newcommand{\be}{\begin{equation}\rm}
\newcommand{\ee}{\end{equation}}
\usepackage{amsmath}

\shorttitle{Measuring CMB Lensing with LDPs}
\shortauthors{Dong et al.}

\begin{document}

\title{Detection of cross-correlation between CMB Lensing and low-density points}

\author[0000-0003-0296-0841]{Fuyu Dong}
\altaffiliation{dongfy2020@kias.re.kr}
\affiliation{School of Physics, Korea Institute for Advanced Study (KIAS), 85 Hoegiro, Dongdaemun-gu, Seoul, 02455, Republic of Korea}
\affiliation{Department of Astronomy, School of Physics and Astronomy, Shanghai Jiao Tong University, Shanghai, 200240, China} 
\author{Pengjie Zhang}
\altaffiliation{zhangpj@sjtu.edu.cn}
\affiliation{Department of Astronomy, School of Physics and Astronomy, Shanghai Jiao Tong University, Shanghai, 200240, China} 
\affiliation{Division of Astronomy and Astrophysics, Tsung-Dao Lee Institute, Shanghai Jiao Tong University, Shanghai, 200240, China}
\affiliation{Key Laboratory for Particle Astrophysics and Cosmology
(MOE)/Shanghai Key Laboratory for Particle Physics and Cosmology,
China} 
\author{Le Zhang}
\altaffiliation{zhangle7@mail.sysu.edu.cn}
\affiliation{School of Physics and Astronomy, Sun Yat-Sen University, 2 Daxue Road, Tangjia, Zhuhai, 519082, P.R. China} 
\author[0000-0002-7336-2796]{Ji Yao}
\affiliation{Department of Astronomy, School of Physics and Astronomy, Shanghai Jiao Tong University, Shanghai, 200240, China}
\author{Zeyang Sun}
\affiliation{Department of Astronomy, School of Physics and Astronomy, Shanghai Jiao Tong University, Shanghai, 200240, China}
\author{Changbom Park}
\affiliation{School of Physics, Korea Institute for Advanced Study (KIAS), 85 Hoegiro, Dongdaemun-gu, Seoul, 02455, Republic of Korea}

\author{Xiaohu Yang}
\affiliation{Department of Astronomy, School of Physics and Astronomy, Shanghai Jiao Tong University, Shanghai, 200240, China} 
\affiliation{Division of Astronomy and Astrophysics, Tsung-Dao Lee Institute, Shanghai Jiao Tong University, Shanghai, 200240, China}
\affiliation{Key Laboratory for Particle Astrophysics and Cosmology
(MOE)/Shanghai Key Laboratory for Particle Physics and Cosmology,
China}

\begin{abstract}
Low Density Points (LDPs, \citet{2019ApJ...874....7D}), obtained by removing high-density regions of observed galaxies, can trace the Large-Scale Structures (LSSs) of the universe. In particular, it offers  an intriguing opportunity to detect weak gravitational lensing from low-density regions.  In this work, we investigate tomographic cross-correlation between Planck CMB lensing maps and LDP-traced LSSs, where LDPs are constructed from the DR8 data release of the DESI legacy imaging survey, with about $10^6$-$10^7$ galaxies. We find that, due to the large sky coverage (20,000 deg$^2$) and large redshift depth ($z\leq 1.2$), a significant detection ($10\sigma$--$30\sigma$)  of the CMB lensing-LDP cross-correlation in all six redshift bins can be achieved, with a total significance of $\sim 53\sigma$ over $
\ell\le1024$.  Moreover, the measurements are in good agreement with a theoretical template constructed from our numerical simulation in the WMAP 9-year $\Lambda$CDM cosmology. A scaling factor for the lensing amplitude $A_{\rm lens}$ is constrained to $A_{\rm lens}=1\pm0.12$ for $z<0.2$, $A_{\rm lens}=1.07\pm0.07$ for $0.2<z<0.4$ and $A_{\rm lens}=1.07\pm0.05$ for $0.4<z<0.6$, with the r-band absolute magnitude cut of $-21.5$ for LDP selection. A variety of tests have been performed to check the detection reliability, against variations in LDP samples and galaxy magnitude cuts, masks, CMB lensing maps, multipole $\ell$ cuts, sky regions, and photo-z bias. We also perform a cross-correlation measurement between CMB lensing and galaxy number density, which is consistent with the CMB lensing-LDP cross-correlation. This work therefore further convincingly demonstrates that LDP is a competitive tracer of LSS.

\end{abstract}


\section{INTRODUCTION}
Due to the gravitational lensing effect, the Large-Scale Structures (LSSs) along the line of sight deflects the Cosmic Microwave Background (CMB) photons trajectories, which can leave measurable imprints on the CMB. The CMB lensing, as a late-time observable, therefore can place competitive constraints on cosmological parameters and help to resolve the Hubble tension~\citep{2020A&A...641A...8P}  between the early- and the late-Universe.


Reconstructed lensing maps contain the integrated information of the overall matter distribution in the universe, and hence such tomography approach allows us to precisely measure the evolution of the growth of structure, as well as increasing the significance of the detection of CMB lensing.  Nowadays, many large and deep redshift surveys, which have broad overlap with the CMB lensing kernel, have made significant progress in finding high-significance detection of cross-correlations.

Over the past few years, many studies have been carried out to investigate the cross-correlation between CMB lensing and LSS tracers, such as galaxy-CMB lensing correlation \citep{2008PhRvD..78d3520H,2011PhRvL.107b1301D,2012ApJ...753L...9B,2012ApJ...756..142V,2012PhRvD..86h3006S,2013ApJ...771L..16H,2013ApJ...776L..41G,2015arXiv150203405O,2015MNRAS.451..849A,2015ApJ...802...64B,2016MNRAS.456.3213G,2018MNRAS.481.1133P,2018PhRvD..98l3526G,2019PhRvD.100d3501O,2020MNRAS.tmp.3681K,2020MNRAS.491...51S,2020JCAP...05..047K,2021MNRAS.501.1481H,2021arXiv210511936H,2021arXiv210511936H,2021arXiv210602551S}, cosmic shear-CMB (polarization) lensing correlation \citep{2012ApJ...759...32V,2015A&A...584A..53K,2018PhRvD..98l3526G,10.1093/mnras/stw1584,Namikawa_2019}, thermal SZ-CMB lensing correlation \citep{2014JCAP...02..030H}. 

Furthermore, contrary to the high-density peaks of the matter distribution, the vast low-density regions (known as cosmic voids) has recently achieved great importance for cosmology.  This is because the voids have many desirable properties for probing the dark energy and can be accurately modeled by the linear perturbation theory, even on small scales. The studies of the correlation between CMB lensing maps with vast low-density regions have been performed recently \citep{10.1093/mnras/stw3299,2017ApJ...836..156C,10.1093/mnras/staa3231,2020ApJ...890..168R,2021arXiv210511936H}, reporting non-zero detection about 3--5$\sigma$.


Similar to the void method, ~\cite{2019ApJ...874....7D} have suggested measuring the stacked shear signals around Low-Density-Points (LDPs), which enables to differentiate several dark energy models.
LDPs are obtained by removing high-density regions of observed galaxies within a certain radius, yielding void-like profiles with negative surface density.
In addition, ~\cite{2021MNRAS.500.3838D} demonstrate that, using a correlation  with CMB temperature maps, the LDP method is a competitive alternative to the existing methods for measuring the integrated Sachs-Wolfe (ISW) effect. In this study, we use the reconstructed convergence map from the Planck observation~\citep{2020A&A...641A...8P} and LDPs from the DESI legacy imaging surveys DR8 to examine the performance of the CMB lensing-LDP cross correlation.


The paper is organized as follows: In Sect.~\ref{sec:data}, we describe the data sets used in our analysis, and explain the methodology  adopted in this study, and then we obtain theoretical predictions based on a numerical N-body simulationin in \S\ref{sec:simu}. In Sect.~\ref{sec:cmp}, we present our results and consistency checks are performed in Sect.~\ref{sec:further}. Our conclusions are given in Sect.~\ref{sec:summary}.

\section{Data analysis}
In this study, we perform the analysis using the photo-$z$ galaxy catalog from the DESI imaging surveys and the Planck CMB lensing maps, in the Galactic coordinate system.

\label{sec:data}
\subsection{Galaxy Catalog}

The photometric galaxy catalog of the DESI imaging surveys is selected upon the three optical bands ($g$, $r$, $z$) and mid-infrared bands observed by the Wide-field Infrared Survey Explorer satellite~\citep{2019ApJS..242....8Z,2016AAS...22831702S,2005IJMPA..20.3121F,2016AAS...22831701B,2018ApJS..239...18A}, for which the DESI Legacy Survey DR8 data covers $\sim20000~\rm deg^2$ in both northern and southern Galactic caps, and source detection uses stacked images of the $(g, r, z)$ bands.

We select galaxy samples by using the DESI DR8 galaxy catalog \citep{2019ApJS..242....8Z}, in which bright stars are excluded and only consider the galaxies with $r < 23.0$ with full five-band photometric measurements.
The photo-$z$ of each galaxy  is estimated by a local linear regression algorithm \citep{2016MNRAS.460.1371B,2018ApJ...862...12G}, based on the nearest neighbors with spectroscopic redshifts, and the redshift error is given as the root-mean-square between the spectroscopic redshifts of these neighbors and the best-fit one. 

\subsection{CMB Lensing Map}
We use the latest PLANCK reconstructed lensing convergence maps and analysis masks provided by~\cite{2020A&A...641A...8P}. In this work our fiducial analysis is based on the CMB lensing maps estimated from the SMICA 
foreground-cleaned CMB maps~\citep{2020A&A...641A...3P} with the temperature-only estimator, where the impact of the thermal Sunyaev-Zeldovich (tSZ) effect has been deprojected using a multi-frequency component separation approach. Moreover, the lensing map determined from both the temperature and polarization maps with the minimum-variance (MV) estimator, is also taken into account for comparison, where the tSZ bias in lensing reconstruction is not removed.  Throughout the paper, these two lensing maps will be referred to as ``tSZ-deproj'' and ``MV'', respectively. 


The spherical harmonic coefficients of the reconstructed lensing convergence maps are provided in the format of HEALPix~\citep{2005ApJ...622..759G}, with $\ell_{\rm max}=4096$, and the associated masks are given as HELAPix maps with the resolution of $N_{\rm side}$=2048. In order to match the resolution of the LDP maps, we downgrade the lensing maps to a lower resolution with $N_{\rm side} =512$ in an appropriate way, to avoid aliasing of power from small to large scales (see details in Appendix A).

Furthermore, noise Planck Full Focal Plane (FFP10\footnote{\url{http://pla.esac.esa.int/pla/\#cosmology}}) simulations~\citep{2020A&A...641A...8P} (300 realizations in total) are used to determine the mean-field bias and the associated statistical error, and specifically, 60 out of the 300 simulations for the mean-field removal and the rest for the covariance matrix estimation. 

\subsection{Survey Masks}
\label{sec:mask}
The mask is generated from the available random catalogs provided by the DR8 website\footnote{\url{https://portal.nersc.gov/cfs/cosmo/data/legacysurvey/dr8/randoms/}}. The random points contain the number of observations in the $g, r, z$ bands, according to sky coordinates drawn independently from the observed distribution.

With $N_{\rm side}=4096$ in the HEALPix scheme, we choose random points which have observations in all $grz$ bands and MASKBITS$=0$ to produce the survey mask, which can populate the same sky coverage and geometry with the galaxy catalog. The objective of the MASKBITS is to provide a general purpose artifact flagging capability. As defined, MASKBITS $\equiv \sum_i 2^{{\rm bit}_i}$, and the bit values, ${\rm {bit}}_i$, which are integers from 0 to 13, indicate the different causes of contamination and are defined specifically for each point of the random catalogues released by DESI. In general, for a given point, the MASKBITS $> 0$ (i.e., ${\rm bit}_i\geq 0$) corresponds to a “bad” one which may potentially suffer contamination that requires flagging. Instead, in our case, since we only consider sources that are not in corrupted pixels, nor near bright stars or globular clusters or large galaxies, i.e., excluding objects with MASKBITS above zero set in the Legacy Surveys catalogs, so that we simply use random points with MASKBITS $=0$ to create a binary mask map that removes all the contaminated sources. In the analysis, all sources lying in the masked regions are excluded.

\subsection{LDP Identification and Over-Density Map}
\label{subsec:LDP-identify}
 \begin{figure}
    \centering
     \subfigure{
     \includegraphics[width=1\linewidth, clip]{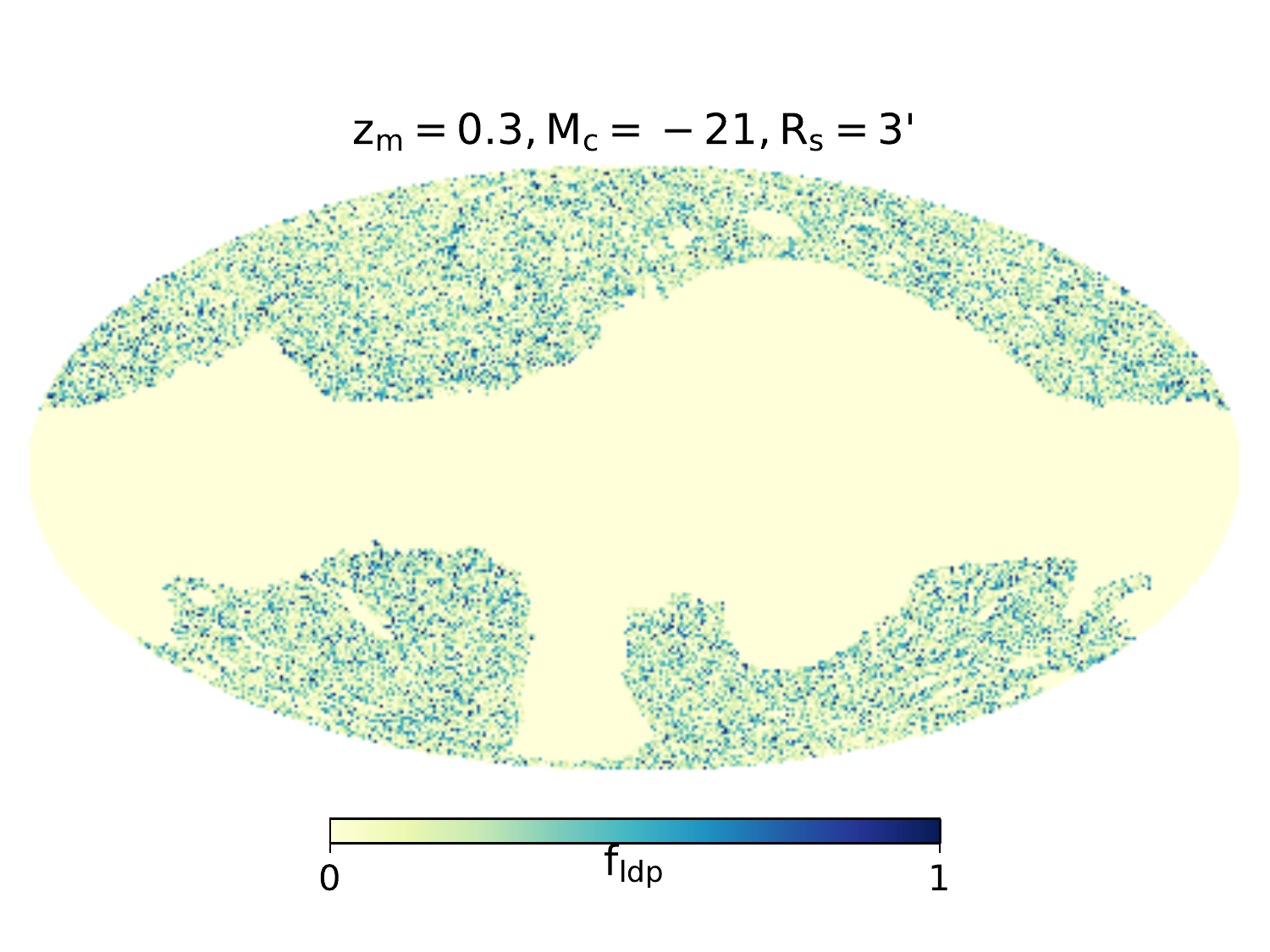}}
     \caption{Map of LDP overdensity, $f_{ldp}$, using HEALPix scheme with resolution $N_{\rm side}$ = 512. A combined mask ($f_{sky}\sim0.44$) from both DESI DR8 catalog  and the Planck survey is applied to the full-sky LDP over-density (i.e., the fraction of low-density points in each pixel). LDPs are generated with a cut radius of $R_s = 3'$ based on the galaxy sample selected with absolute magnitude cut of $M_c= -21$ and photo-$z$ cut of $0.2<z < 0.4$.}
    \label{fig:fldp-map}
\end{figure}


\begin{table*}[!htp]
    \footnotesize
    \centering
    \caption{Galaxies that are brighter than $M_c$ in the DR8 photo-$z$ catalogue.}
    \label{table:DR8}
\begin{tabular}{cccccccccc}
\hline
& && $n_{\rm gal}\,(\times 10^6)$ & &&& $\overline{f}_{\rm LDP}$\\
\hline
 {$z_m\setminus M_c$} &  -20.5 &-21 &-21.5 & -22 && -20.5 &-21 &-21.5 & -22 \\
\hline
0.1 &1.73   &0.81   &0.28      &-       && 0.55  & 0.75   & 0.9     &-  \\
0.3 &-       &4.53   &1.31       &0.26 && -        &0.22 &0.62     &0.9 \\

0.5 &-      &6.52         & {1.66}       & {0.24}   && -        &  {0.16}        &   {0.55}   &   {0.91}   \\
0.7 &-      &  {6.98}         &  {1.66}       &  {0.21}   && -        &   {0.14}        &   {0.56}   &   {0.91}  \\

0.9 &-      &  {5.81}         &  {1.11}       &  {0.14}   && -        &   {0.16}        &   {0.64}   &   {0.94}   \\
1.1 &-      &  {3.16}         &  {0.54}       &  {-}   && -        &   {0.31}        &   {0.8}   &   {-}   \\
\hline
\multicolumn{8}{l}{{$z_m$}: the redshift slice centered at $z_m$ with bin-width of 0.1;}&\\
\multicolumn{8}{l}{{$n_{\rm gal}$}: the number of galaxies in each galaxy sample;}&\\
\multicolumn{8}{l}{{$\overline{f}_{\rm LDP}$}: the average value of $f_{\rm LDP}$,  identified with   $R_s=3'$.}&\\
\end{tabular}
\end{table*}

The LDPs are certainly dependent on galaxy samples. Using the procedure developed in~\cite{2021MNRAS.500.3838D}, we first use the $r$-band absolute magnitudes\footnote{The K-correction is not applied to the absolute magnitudes since the absence of K-correction is not an issue for the LDP generation. This is because galaxies within the same photo-$z$ bin have similar K-corrections and the LDPs are only sensitive to the relative brightness of these galaxies.} to select galaxies that are brighter than a certain absolute magnitude, $M_c$, and within a redshift slice of $z_m\pm\Delta z$ with $\Delta z = 0.1$ which is chosen due to the photo-$z$ error dispersion $\lesssim 0.1$ on average.

The r-band absolute magnitude can be expressed in terms of    
\begin{equation}
M_r-5\lg h = m_r - 5\lg\left(\frac{d_L}{h^{-1}\rm Mpc}\right)-25\,,
\end{equation}
where $d_L$ is the luminosity distance determined by photometric redshift, $h$ is the dimensionless Hubble parameter. In the following, we use a parameter $M_c$ for magnitude cut, which represents the absolute magnitude limit on the faint end. In this galaxy sample, we surround each galaxy with an angular radius of $R_s$ and remove all pixels lying within this radius from the sky. The remaining regions are defined as ``LDP'' candidates.

We sample LDPs on equal-area high-resolution HEALPix grids at $N_{\rm side} = 4096$, corresponding to an angular resolution of 0.859'.
According to \cite{2021MNRAS.500.3838D}, LDPs physically correspond to underdense regions, and the generation of LDPs in principle depends on the choice of the cut radius $R_s$ to each galaxy. Statistically, a larger $R_s$ would lead to a more negative density contrast $\delta_m$. After generating LDPs, we apply the same procedure to produce the pixelized LDP over-density field as in~\cite{2021MNRAS.500.3838D}, to which we refer readers for more details, so that,
\begin{equation}
\label{eq:fldp}
\delta_{l}=\frac{f_{\rm LDP}-\overline{f}_{\rm LDP}}{\overline{f}_{\rm LDP}},\,
\end{equation}
where $f_{\rm LDP}$ is the proportion of area occupied by LDPs in a coarse cell. For a given LDP sample, the larger $f_{ldp}$ is, the more negative $\delta_l$ becomes in each cell.

We make a choice on the coarse cells being used in order to reduce the impact of mask and edge effect on our calculations on $\delta_l$. We require that the random points satisfy the selection criteria in     \S\ref{sec:mask} should fill more than $70\%$ of the area of each selected cell.

Moreover, based on the fact that, fewer LDPs will be identified by selecting more galaxies, and the number of galaxy samples varies with redshift for a given projected number density, we have to control the number of galaxy samples such that the LDP distribution become neither too populated nor too sparse, making the statistics of $\delta_{\rm LDP}$ accurate. In practice, for photo-$z$ bin of $0.01\le z\le0.2$, we choose $M_c$ to be -20.5, -21, and -21.5, respectively. For $0.3\le z\le0.9$, $M_c$ is set to be -21, -21.5, and -22, and we then set $M_c$ to be -21.5 and -22 for $1.0\le z\le1.2$

In Tab.~\ref{table:sn}, we summarize the galaxy samples and the corresponding average density of LDP, by varying galaxy selection criterion. In Fig.~\ref{fig:fldp-map}, we show one example of the $f_{\rm LDP}$ distribution, generated with the following parameters, $z_m = 0.3, M_c= -21$, and $R_s = 3'$. Both the galaxy and CMB masks have been applied to the full-sky LDP map, corresponding to a sky coverage of $f_{sky}\simeq 0.44$.

\section{Theoretical prediction from Numerical Simulation}
\label{sec:simu}
From a theoretical perspective, due to the complicated observational effects, and the non-trival relationship between LDP and galaxy samples, one requires simulations to validate and interpret the final cross-correlation signal, although implementing LDP analysis directly to observed data is straightforward. We will use an N-body simulation to generate mock data including galaxy catalogs and LDPs as well as lensing convergence maps, under a given observational condition, which are then used to predict the CMB lensing-LDP cross-correlation signal.

\subsection{Simulation Data}
\begin{figure}
    \centering
    \subfigure{
     \includegraphics[width=1\linewidth, clip]{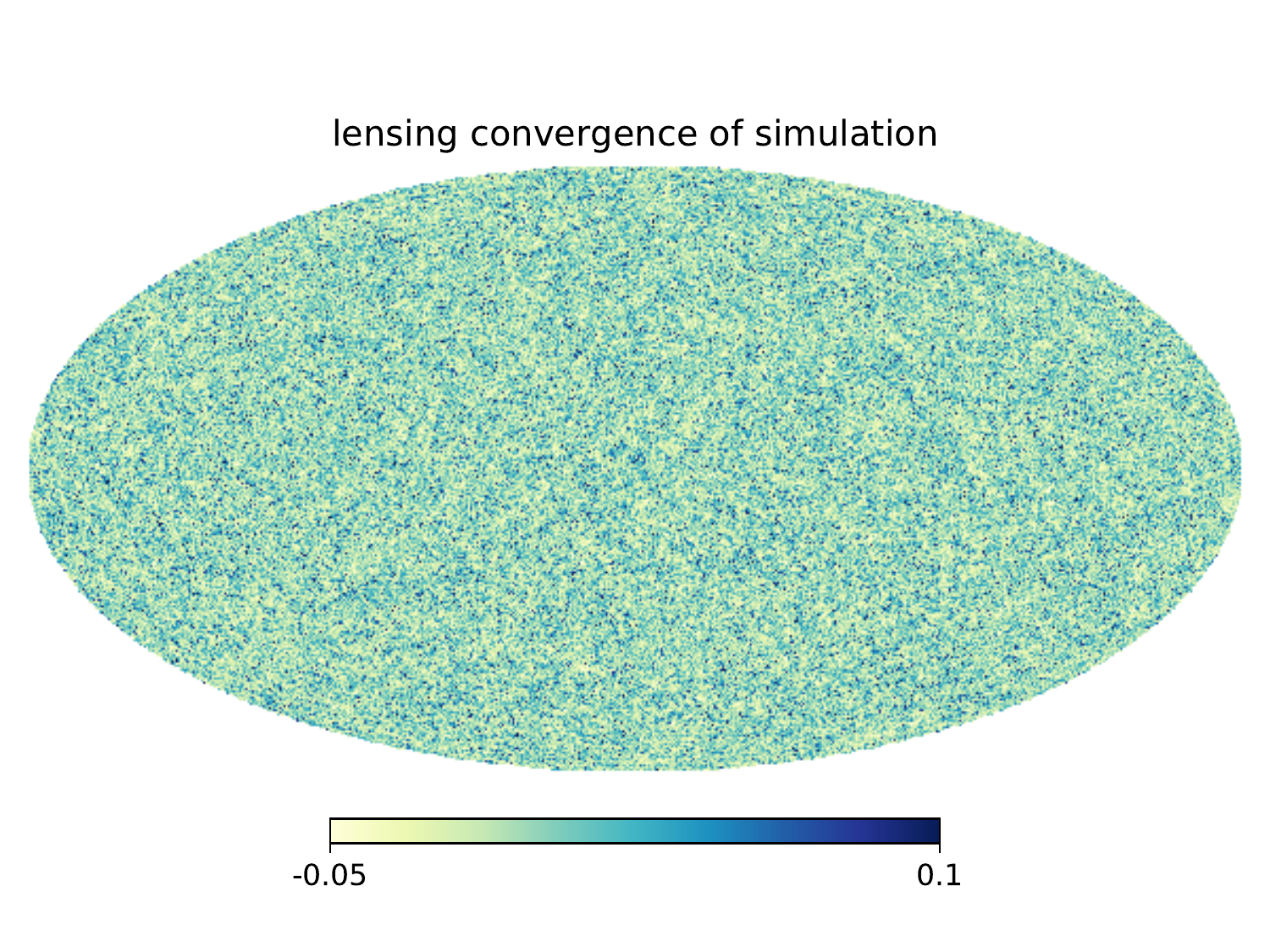}}
     \subfigure{
     \includegraphics[width=1\linewidth, clip]{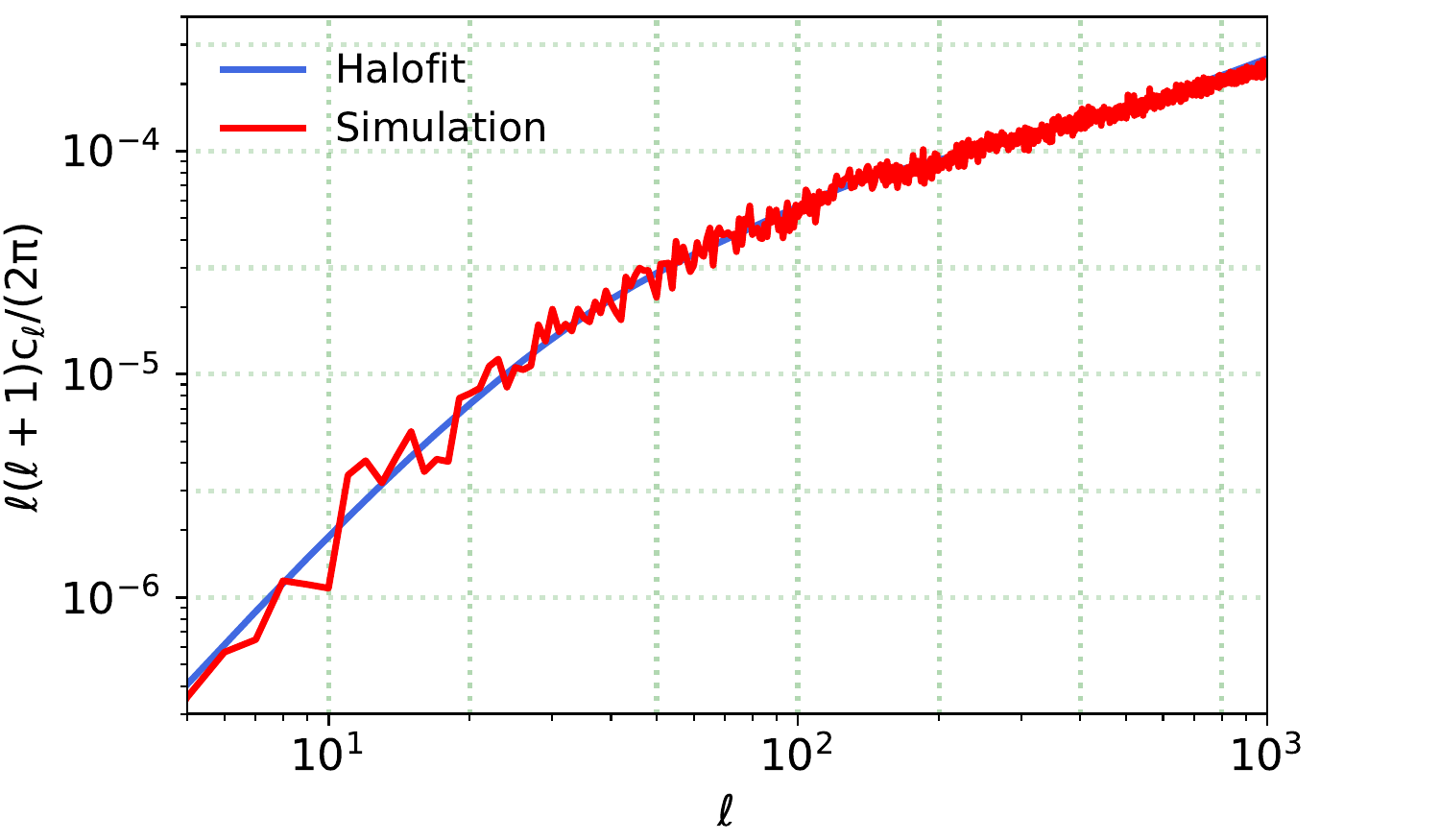}}
     \caption{CMB lensing convergence map (upper panel) and corresponding power spectrum (lower panel), generated from our N-body simulation by stacking density field along LOS in mock light-cone for the redshift bins $z<0.972$.  The theoretical prediction (blue) computed in Limber approximation within the same redshift range and using the Halofit nonlinear power spectrum is shown as blue line, showing a good consistency between the simulated and the theoretical results.}
    \label{fig:kappa-map}
\end{figure}

To accurately model the realistic LSS matter distribution and the CMB lensing field, we use a high-resolution N-body simulation from the Cosmic-Growth simulation series~\citep{2019SCPMA..6219511J}, which uses a  flat $\Lambda$CDM cosmology  with parameters: $\Omega_m=0.268,~\Omega_{\Lambda}=0.732,~\sigma_8=0.85,~h=0.71,~n_s = 0.968$, with $3072^3$ dark matter particles in a box size of 1200 ${\rm Mpc}/h$. Halos in simulation are identified with FoF group finder~\citep{1976ApJS...32..409T,1982ApJ...257..423H,1985ApJ...292..371D}, and subhalos with Hierachical Bound-Tracing (HBT) method~\citep{2012MNRAS.427.1651H}.


\subsection{Generation of LDPs with mock galaxies}
\label{sub:mock}
Similar to the procedure in~\citep{2021MNRAS.500.3838D}, the mock photo-$z$ catalog is generated based on the HBT subhalo catalog. First, we draw correspondence between galaxies and halos/subhalos through matching the galaxy-subhalo abundance (SHAM). And then, subhalos from different snapshots are used to fill the corresponding radial distance slices. Finally,  the mask of DR8 is applied in the simulation to ensure the same angular selection of galaxies as in the observation. In addition,  we also generate scatters in the redshifts and luminosities of mock galaxies so as to better mimic the real observation.

In observation, galaxies are provided with photo-z errors, according to which we disturb the redshift of mock galaxies. By assuming a Gaussian distributed photo-$z$ error with zero mean and a scatter of $\delta_z$, we randomly shift the galaxy redshifts and adjust the absolute magnitudes of mock galaxies accordingly based on $z$-$z_{\rm photo}$. Moreover, to mimic the galaxy-halo/subhalo relation, a constant scatter $\sigma_{\rm Mag} = 0.375$ dex to each galaxy is taken into account~\citep{2008ApJ...676..248Y}. After taking these uncertainties into account, we re-performe the SHAM to guarantee that the luminous function of the simulated galaxies is the same as the observed one. By doing so, we generate a simulated photo-$z$ catalog in the simulation with $z<0.8$. 

Note that in the above procedures, subhaloes in the redshift bin of $0.8\leq z<1$ are also used to generate realistic volume-limited mock galaxies in lower redshift bins, since there is a photo-z induced leakage mixing galaxies between higher and lower redshift bins.

\subsection{Construction of full-sky lensing convergence map}

Under the Born approximation~\citep{2002ApJ...574...19C}, the convergence $\kappa$ for CMB photons at the last-scattering surface with $z_{s}\simeq 1100$) and direction $\widehat{n}$ is the projected density fluctuation $\delta_m(\hat{n},z)$ weighted by the lensing kernel along LOS, which reads  
\begin{equation}\label{eq:k}
\kappa(\hat{n})=\int_0^{\chi_s}\delta_m(\hat{n},z)W^\kappa(z,z_s)d\chi\,,
\end{equation}
where $\chi$ is the comoving distance to redshift $z$, and $W^\kappa$ is the lensing kernel expressed by
\begin{equation}
W^\kappa(z,z_s) = \frac{3}{2}\left(\frac{H_0}{c}\right)^2\Omega_m(1+z)\chi(z)\left[1-\frac{\chi(z)}{\chi(z_s)}\right]\,.
\end{equation}
We could construct the map of lensing convergence by stacking projected surface mass densities along the LOS.

\cite{2018ApJ...853...25W} has confirmed a good agreement among the predicted angular power spectrum of convergence from the Born approximation by stacking density field along LOS in mock light-cone and that from the full-sky ray-tracing simulation at  $\ell \lesssim 4000$, with relative deviation $< 10\%$. In this study, we focus on the multipoles $\ell<1024$ so that the Born-approximation is accurate enough for this purpose.


To a reasonable approximation, Eq.~\ref{eq:k} can be written in terms of:
\begin{equation}
\kappa(\hat{n})\simeq\sum_i\delta^\Sigma_i(\hat{n})W_i\Delta\chi_i\,,
\end{equation}
where $\delta^\Sigma_i$ is the surface overdensity of the $i$-th
redshift bin ($z_i-\Delta z_i<z<z_i+\Delta z_i$), and $W_i\simeq W(\chi_i,\chi_s)\Delta \chi_i$ and $\Delta \chi_i=\chi(z_i+\Delta z_i/2)-\chi(z_i-\Delta z_i/2)$. Such approximation is reasonably good as long as $\Delta z_i< 0.2$~\citep{2016PhRvD..94h3520Y}.

We construct the full-sky past light-cone of the observer  in the redshift range of $0\leq z \leq 0.97$. The light-cone is built based on the nine output snapshots at redshifts of 0, 0.058, 0.151, 0.253, 0.364, 0.485, 0.616, 0.76 and 0.916, respectively, corresponding to spherical shells of comoving thickness $\Delta \chi \simeq 300 {\rm Mpc}/h$.  This essentially guarantees that the lensing convergence maps generated by mock samples within the redshift range of $z<0.8$. The particles located inside each of these snapshot are projected onto spherical shells of surface over-density $\delta^\Sigma_i$ of the $i$-th redshift shell in $[(z_{i-1}+z_{i})/2,(z_i+z_{i+1})/2]$, sampled on an HEALPix grid at $N_{\rm SIDE} = 1024$ (angular size 3.44').  During this process, we assume periodic boundary conditions in order to avoid discontinuities of matter distribution.

In Fig.~\ref{fig:kappa-map}, the upper panel shows the simulated CMB convergence map based on our mock samples and, to verify the simulation results in terms of the theoretical predictions, the lower one shows the measured angular power spectrum of such convergence map is in good agreement with the theoretical prediction. Theoretically, the CMB lensing auto-correlation in the Limber approximation\citep{1953ApJ...117..134L} is
\begin{equation}
\label{eq-clkk}
C_l^{\kappa\kappa}=\int_0^{\chi_s} d\chi\frac{W^\kappa(\chi)^2}{\chi^2}P_{mm}\left(k=\frac{l+1/2}{\chi},z\right)\,,
\end{equation}
where $W^\kappa(\chi)$ is the CMB lensing kernal, $P_{mm}(z)$ is the matter power spectrum evaluated at wavenumber $k$ at redshift $z$, and one can compute it using the Halofit dark matter nonlinear power spectrum ~\citep{2003MNRAS.341.1311S}.  For the theoretical prediction curve shown in Fig.~\ref{fig:kappa-map}, we adopt the same redshift range as the simulation, $z<0.972$.

The angular cross-spectrum between LDP overdensity field and CMB lensing convergence can be similarly written as
\begin{equation}
\label{eq-cllk}
C_\ell^{l\kappa}=\int_0^{\chi_s} d\chi\frac{W^l(\chi)W^\kappa(\chi)}{\chi^2}P_{lm}\left(k=\frac{l+1/2}{\chi},z\right)\,,
\end{equation}
where $W^l(\chi)$ is the projection kernel for LDPs, and $P_{lm}$ represents the  cross-power spectrum of matter and LDPs. However, instead of modelling $W^l(\chi)$ and $P_{lm}$ analytically, the theoretical prediction for $C_\ell^{l\kappa}$ is estimated by using our simulations.

\section{Results}
\label{sec:cmp}
Using the data analysis pipeline and the simulated data described above, we now begin to perform tomographic cross-correlation measurements between the Planck CMB lensing maps and the LDPs, constructed from the DESI legacy imaging surveys DR8,  and present their theoretical interpretations.

\subsection{Lensing of the CMB detected by LDPs in Observation}

\begin{figure*}
    \centering
    \subfigure{
     \includegraphics[width=1\linewidth, clip]{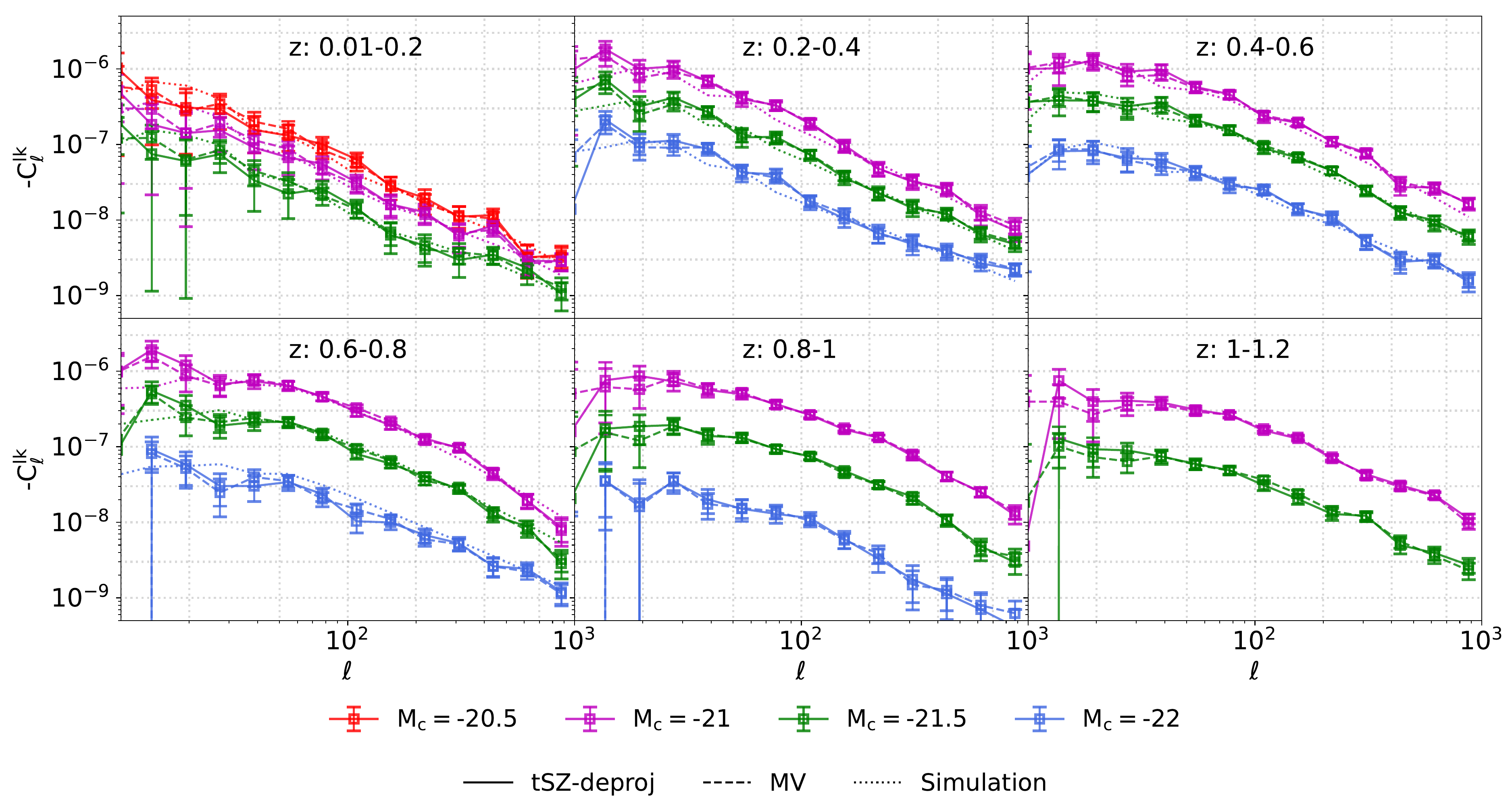}}
    \caption{
    The cross-correlation between CMB lensing and LDPs  measured for six redshift bins: $z_m=$0.1, 0.3, 0.5, 0.7, 0.9 and 1.1 with $R_s=3'$. All the signals are multiplied by minus one. The results for LDPs generated with different parameters are distinguished in different colored lines.  There are two different lines shown for each color. The solid line shows signal measured with tSZ-deproj map in observation, the dashed line shows signal measured with MV map in observation, and the dotted line shows signal measured from the simulation.}
    \label{fig:power-obs}
\end{figure*}

Using the photo-$z$ samples of the DESI DR8 galaxy catalog, in this study we split the galaxy sample into 6 redshift bins: [0.01, 0.2], [0.2, 0.4], [0.4, 0.6], [0.6, 0.8], [0.8,1] and [1,1.2] , as shown in Tab.~\ref{table:sn}. 
In the following analysis, we choose a binning scheme with 13 logarithmically-spaced bins from $\ell_{\rm min}=12$ up to $\ell_{\rm max}=1024$,  and the averaged amplitude of power spectrum in each $\ell$-bin with $N_\ell$ being the number of multipoles is estimated in terms of  
\begin{equation}
\overline{C}_\ell=\frac{1}{N_\ell}\sum_{\ell'=l-N_\ell/2}^{\ell+N_\ell/2}C_{\ell'}\,,
\end{equation}
where $\ell$ is the mean multipole value for each $\ell$-bin, and $C_{\ell'}$  measured with the healpy software package \citep{Zonca2019,2005ApJ...622..759G}.
Also note that, the band powers are essentially uncorrelated if we set the bin widths to be sufficiently large. In our case for $f_{sky}\sim 0.44$, we have the widths $\Delta \ell \gg f^{-1/2}_{sky}$, so that the band powers and associated statistical errors between adjacent bins will be essentially uncorrelated since the mode coupling matrix will be fairly diagonal.

A tomographic cross-correlation measurement between LDPs at the 6 redshift bins and CMB lensing convergence maps from \cite{2020A&A...641A...8P} is shown in Fig.~\ref{fig:power-obs} (note the minus sign added in $C_\ell^{\kappa l}$), where the LDPs are produced with galaxies with $R_s=3'$. As shown, (i) the negative correlations indicate that the underlying matter densities $\delta_m$ at LDPs are statistically negative and hence the LDP field indeed probes underdense regions as expected; (ii) the cross-correlation signals by using  the ``MV'' CMB convergence map and the ``tSZ-deproj'' map are well consisitent with each other, clearly lying within 1-$\sigma$ of one another with a mean difference of about $0.99\pm 0.03$ in the amplitude averaged over $\ell\ge17$ and $1\pm0.07$ over $\ell\ge91$; (iii) the error bars on the cross-spectra are measured with the Planck noise simulations, and the errors from ``MV'' map  are slightly smaller than those from ``tSZ-deproj'' as the ``MV'' map constructed by all temperature and polarization modes rather than the temperature-only ``tSZ-deproj'' map. 

\subsection{Covariance Matrix and Detection Significance in Observation}
\begin{figure}
    \centering
    \subfigure{
     \includegraphics[width=1\linewidth, clip]{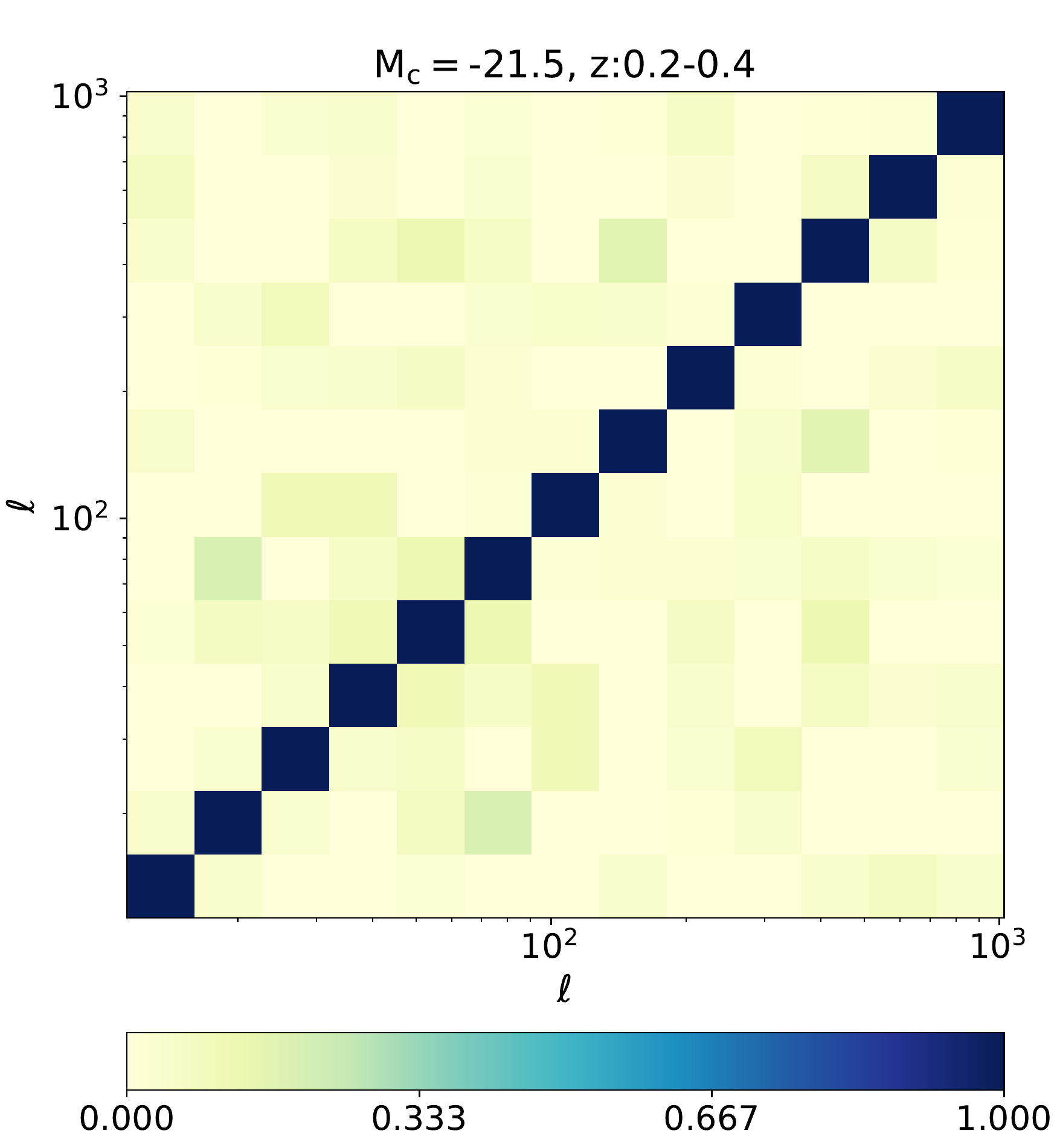}}
   \caption{Example of the normalized covariance matrix for the LDP-CMB lensing cross-correlation with parameters of  $M_c=-21.5$ and $z_m=0.3$, estimated from 240 simulated realizations of Planck convergence maps.}
    \label{fig:cov-21.5}
\end{figure}

\begin{table*}
    \footnotesize
    \centering
    \caption{S/N of $C_{\ell}^{\kappa l}$ derived from the observational measurements over the multipole range of $12\leq \ell \leq 1024$, with $R_s=3'$.}
    \label{table:sn}
\begin{threeparttable} 
\begin{tabular*}{\textwidth}{@{\extracolsep{\fill}\quad}cccccccccccc} 
\hline
$\kappa$ map & \diagbox[innerwidth=4.5cm,height=1.2cm,innerleftsep=-26pt]{$z_m$}{${\rm{S/N}_{null(model)}}$\tnote{a}}{$M_c$} && -20.5 &-21 &-21.5 &-22 &Max(${\rm {S/N}^{ tot}_{null}}$)\tnote{b}\\
\hline
\multirow{6}{*}{"tSZ-deproj"}  &0.1 && 10.7 (10.1)& 10.6 (10.) & 9.2 (8.8) & -   &\multirow{6}{*}{53.3}\\
&0.3 && - & 16.6 (16.) & 16.8 (16.3) & 15.7 (15.1)      \\
&0.5 && - & 21.2 (20.7) & 21.4 (21.1) & 16.5 (16.1)          \\ 
&0.7 && - & 21.4 (21.) & 22.3 (22.) & 13.5 (13.1)         \\ 
&0.9 && - & 25.5 \;\,(\;-\;) & 23.8 \;\,(\;-\;) & 10.1 \;\,(\;-\;)         \\ 
&1.1 && - & 28.9 \;\,(\;-\;) & 17.7 \;\,(\;-\;) & -          \\ 
\hline
\multirow{6}{*}{"MV"} &0.1 && 12.2 & 12.3& 11.3 & -   &\multirow{6}{*}{60.6}  \\
&0.3 && - & 19.4& 19.7 & 17.3      \\
&0.5 && - & 23.8 & 23.3 & 18.4         \\ 
&0.7 && - & 24.5 & 24.7 & 15.5          \\ 
&0.9 && - & 31. & 27.1 & 12.          \\ 
&1.1 && - & 31.6 & 21.2 & -         \\ 
\hline
\end{tabular*}%
\begin{tablenotes}
\item[a] ${\rm {S/N}_{null}} = (\chi^2_{\rm null})^{1/2}$ and ${\rm (S/N)_{model}}= (\chi^2_{\rm null}-\chi^2_{\rm min})^{1/2}$, defined in Eqs.~\ref{eq:sn},~\ref{eq:snr}.
\item[b] The maximum ${\rm (S/N)^{tot}_{null}}$ achieved by combining the respective maximum ${\rm(S/N)_{null}}$ of each bin (i.e., choosing appropriate $M_c$ listed here).  
\end{tablenotes}
\end{threeparttable}

\end{table*}

We cross-correlate the LDP maps with the publicly available Planck FFP10 simulations to empirically estimate the covariance matrix for the cross-power spectrum through, $C_{ij}=\langle C_{\ell_i}C_{\ell_j}\rangle-\langle C_{\ell_i}\rangle\langle C_{\ell_j}\rangle$ with $i$($j$)=$1,\dots,n$, 
where $\ell_i$ denotes the $i$-th bandpower bin, $n$ the number of total bandpower bin, and the ensemble average is taken over 240 realizations of CMB lensing and LDP maps.

One example of the normalized covariance matrix ($C_{ij}/\sqrt{C_{ii}C_{jj}} $), with parameters of  $z_m=0.3$ and $M_c=-21.5$, is shown in Fig.~\ref{fig:cov-21.5}, clearly indicating that the off-diagonal components of the covariance matrix are almost negligible, $<20\%$ of the diagonal ones. The 1-$\sigma$ error for each $\ell$-bin shown in Fig.~\ref{fig:power-obs} can thus be well approximated by $C_{ii}^{1/2}$.  For the purpose of accuracy, we, however, still use the full covariance matrices for all calculations in our results 


To evalute the significance of the cross-correlation detection, we use the $\chi^2$ statistic. For LDPs at a given photo-z bin $z_i$, the cumulative $\chi^2$ for the null hypothesis, $\chi^2_{\rm null}$, is calculated by running over all  $\ell$-bins (denoted by the indices $i,j$) by following 
 \begin{equation}
 \label{eq:sn}
    \chi^2_{\rm null} (z_i) = \boldsymbol{d^T C^{-1} d} =  \sum_{i,j}C^{\kappa l}_{\ell_i,\rm data}C_{ij}^{-1}C^{\kappa l}_{\ell_j,\rm data}\,,
\end{equation}
where ${\boldsymbol d}$ denotes the data vector and $C^{-1}$ is the inverse of the covariance matrix described above.  Due to the fact that LDP sampes of different photo-z bins are almost uncorrelated, the total $\chi^2_{\rm null}$ is calculated by summing over the entire 6 photo-$z$ bins ($z_i =1,2,\dots, 6$). The resulting $\rm (S/N)_{null}$ then read, 
\begin{equation}
 {\rm (S/N)^{tot}_{null}} = (\sum_{z_i}{\rm (S/N)}^2_{\rm null}(z_i))^{1/2} = \left(\sum_{z_i} \chi^2_{\rm null} (z_i)\right)^{1/2}\,.
\end{equation}
Furthermore, the significance of the cross-correlation detection for the preference of the best-fit model over the null hypothesis can be simply defined as,
\begin{equation}
  \label{eq:snr}
  \rm ({S/N})_{model} =  \sqrt{\chi^2_{\rm null} - \chi^2_{\rm min}}\,.
\end{equation}  
Here $\chi^2_{\rm min}$ corresponds to the minimum $\chi^2$ for the best-fit theoretical model for the cross signal, which is calculated by 
\begin{eqnarray}
 \label{eq:chi2min}
  \chi^2&=&  \sum_{i,j} \delta C_{\ell_i}^{\kappa l} C_{ij}^{-1} \delta C_{\ell_j}^{\kappa l} \,, \\ 
  {\rm with\quad}  \delta C_{\ell_i}^{\kappa l} &\equiv&  C_{\ell_i,\rm data}^{\kappa l}-A_{\rm lens}C_{\ell_{i,\rm sim}}^{\kappa l}\nonumber\,\,,
\end{eqnarray}
where the theoretical model has been chosen as $A_{\rm lens}C_{\ell_i,\rm sim}^{\kappa l}$, a template $C_{\ell_i,\rm sim}^{\kappa l}$ measured from our simulations multiplied by a free scaling factor $A_{\rm lens}$ for the lensing amplitude. The best-fit $A_{\rm lens}$ can be determined by minimizing the quantity $\chi^2$.

The S/N values for the various maps and LDP catalogs are summarized in Tab.~\ref{table:sn}. We find that the $\rm (S/N)_{null}$ in each photo-$z$ bin becomes different according to different choices of magnitude cut in sample selection and CMB lensing maps (``MV'' or ``tSZ-deproj''). For example, when using the ``tSZ-deproj'' map, the ${\rm (S/N)_{null}}$ for $z$-bin $0.4 \leq z \leq0.6$ will vary from 22.1 to 16.6 for $M_c =$ -21 and -22, respectively.  Interestingly, a higher redshift bin usually yields a higher detection significance, which is due to the fact that CMB lensing efficiency will reach the maximum at $\sim 2$. Finally, the total S/N from all the 6 photo-$z$ bins can be achieved, ${\rm (S/N)^{tot}_{null}} = 53.3, 60.6$, for the ``tSZ-deproj'' and ``MV'' cases, respectively. These high signal-to-noise measurements strongly validate the LDP as an effective tracer of LSS. 

\subsection{Comparison with the N-body Simulation}
\begin{figure*}
    \centering
    \subfigure{
     \includegraphics[width=1\linewidth, clip]{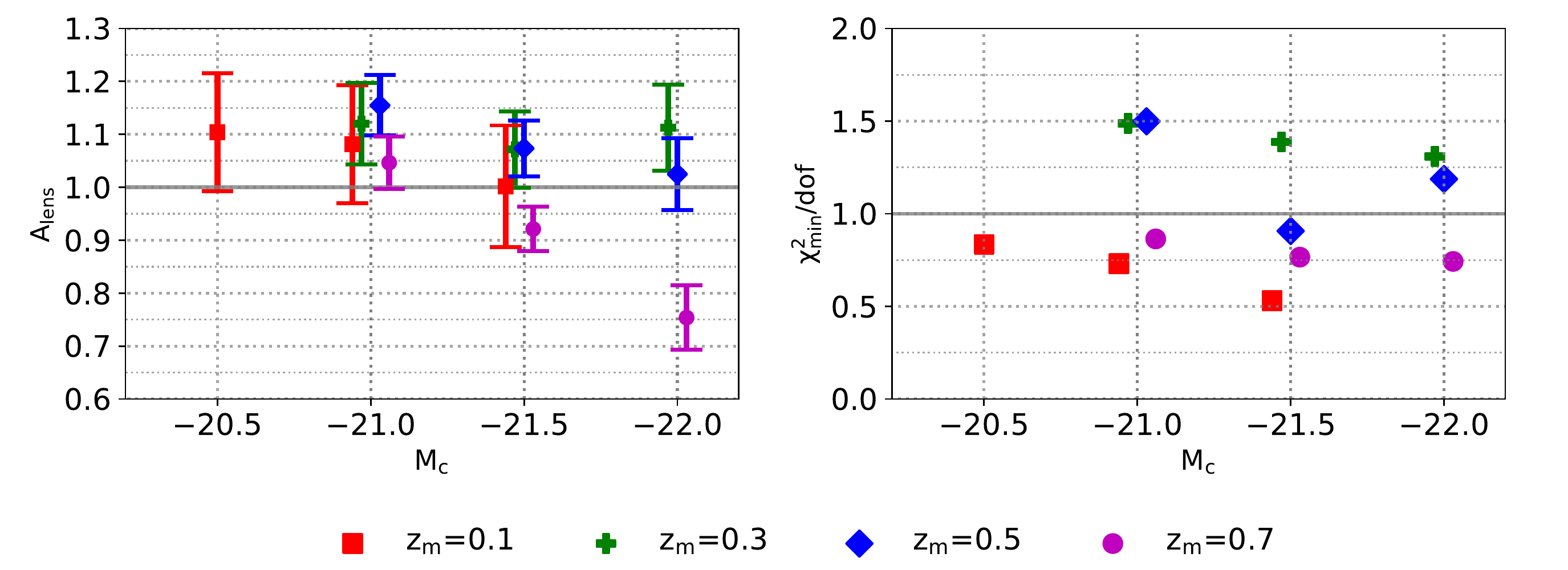}}
   \caption{Left panel: dependence of the estimate of $A_{\rm lens}$ derived from the observed cross-spectrum $C_{\ell}^{\kappa l}$ on the magnitude cut $M_c$ and the photo-$z$ bin. Right panel: the corresponding reduced $\chi^2$, with $dof = 11$.} 
    \label{fig:simu-obs-ldp}
\end{figure*}

In the above we show tomography CMB lensing measurements spanning the redshift range of $0.01<z<1.2$. But in view of the quality of redshift measurement of galaxies and reliable construction of mock samples, we define sources with $z\leq$ 0.6 as our {\it baseline samples}.

Using our simulated both the CMB convergence and LDP maps, described in \S\ref{sec:simu}, we calculate cross-power spectra of the CMB-LDP cross-correlation as a theoretical prediction. These results are shown as dotted lines in Fig.~\ref{fig:power-obs}. For a given photo-$z$ bin, the best-fit amplitude, $A_{\rm lens}$, can be determined by minimizing the $\chi^2$, as defined in Eq.~\ref{eq:chi2min}, where we consider the multipole range of $22\leq \ell \leq 724$ for fitting.

In the ``tSZ-deproj'' case, the best-fit amplitudes and the corresponding reduced chi-square $\chi^2_{\rm min}/dof$ are shown in Fig.~\ref{fig:simu-obs-ldp}, where the LDP maps are generated from the DESI DR8 catalog. The best-fit values of $A_{\rm lens}$ are mostly consistent with the theoretical prediction, within 2-$\sigma$, thus providing a good fit to the data. One may notice that the amplitude from the high redshift bin ($z_m=0.7$) with magnitude cuts of $M_c=22$ will be underestimated, significantly deviating from the simulation result with about 4-$\sigma$, which might be due to potentially large photo-$z$ errors in the photometric measurements. More detailed discussions about photo-$z$ errors as well as various sysmetiactic effects are present in Sect.~\ref{sec:further}. 

\begin{table}
    \footnotesize
    \centering
    \caption{Best-fit $A_{\rm lens}$ and associated $1$-$\sigma$ uncertainty as well as the reduced $\chi^2_{\rm min}$ with $dof=32$, estimated from different combinations of LDP baseline samples from the three $z$-bins with  $z_m=0.1,0.3,0.5$ and using Planck "tSZ-deproj" lensing map.}
    \label{table:alens}
\begin{threeparttable}    
\begin{tabular}{cccc}
\hline
 Sample &Max($M_c$)\tnote{a} &-21.5 &Min($M_c$)\tnote{b} \\
\hline
$A_{\rm lens}$ & 1.14$\pm$0.04     & 1.06$\pm$0.04     &1.05$\pm$0.05\\
$\chi^2_{\rm min}/dof$ & 1.27 & 0.94 & 1.01   \\
\hline
\end{tabular}
\begin{tablenotes}
\item[a] Max($M_c$) refers to selecting fainter but more galaxies, by setting $M_c= -20.5, -21, -21$ for the three $z$-bins, respectively.
\item[b] Min($M_c$) refers to selecting brighter but fewer galaxies, by setting $M_c= -21.5, -22, -22$ for the $z$-bins, respectively.
\end{tablenotes}
\end{threeparttable}
\end{table}

To fully exploit the the measured the cross-power spectra at various redshifts, we combine the spectra at each redshift bin to make a total constraint on $A_{\rm lens}$. Since the estimate of $A_{\rm lens}$ from a given redshift also depends on the choice of LDP samples as well, we summarize the total constraints on $A_{\rm lens}$ based on different magnitude cuts, presented in Tab.~\ref{table:alens}. As seen, if using LDPs generated from a relatively larger number but fainter galaxies (as referred to "Max($M_c$)") by setting $M_c =$ -20.5, -21 and -21 for $z_m=$ 0.1, 0.3, 0.5, respevtively, we obtain the overall constraint of $A_{\rm lens}=1.14\pm0.04$ by combining all signals in the redshift range of $0\leq z\leq 0.6$. Alternatively, if considering LDPs from relatively brighter but smaller number of galaxy samples (referred to "Min($M_c$)") where $M_c=$ -21.5, -22, -22, respectively, for these three bins, the derived $A_{\rm lens}$ is $1.05\pm0.05$, about 8\% lower than the "Max($M_c$)" case. Furthermore, if fixing the magnitude cut to $M_c=-21.5$ for all redshift bins, the similar constraint to the case "Min($M_c$)" is obtained, $A_{\rm lens} = 1.06\pm0.04$. In all these three cases, the corresponding minimum reduced $\chi^2$, are $\chi^2_{\rm min}/dof =$ 1.27, 0.94 and 1.01, with $dof =32$ for the cases of "Max($M_c$)", "$M_c$ fixed" and "Min($M_c$)", respectively, so that the the best-fit model is consistent with data. To make this statement more quantitatively, we calculate the probability-to-exceed (PTE) of the best-fit model. Using $\chi^2_{\rm min}=$ 40.6, 30.1 and 32.3 for $dof =32$, we find PTE $= 0.1415$ for "Max($M_c$)", 0.56 for "$M_c$ fixed", and 0.45 for "Min($M_c$)". Therefore, we find no strong evidence of potential systematics and the model thus provides a good fit to the data.


From Fig.~\ref{fig:power-obs}, we see that, although the measured $C_\ell^{\kappa l}$ from different LDP samples can vary significantly in amplitude, even by an order of magnitude, the good agreement on $A_{\rm lens}$ derived from the observation and from the simulation clearly indicates a robust detection on the lensing-LDP signal. Thus, the LDP method provides a robust test on the $\Lambda$CDM model.

In addition, ~\cite{2021MNRAS.501.1481H} has reported the overall scaling in amplitude, $A_{\rm lens}$, to be $0.901 \pm 0.026$, by using a cross-correlation between Planck lensing convergence maps and DESI LRG samples, which appears to be lower than our results. This is probably because they utilized a different sample of galaxies to us. To cross-check, we also conduct a cross-correlation measurement by using galaxies instead of LDPs, and find the derived $A_{\rm lens}=1\pm0.04$ for our baseline samples and $M_c=-21.5$, as discussed in Appendix B.

\section{Consistency Checks}
\label{sec:further}
In this section, we will present several consistency checks carried out to ensure that our measurements are accurate and robust to the data selection and observational effects. 

\subsection{Hemispherical Asymmetry}

\begin{table*}[!htb]
    \footnotesize
    \centering
    \caption{Best-fit $A_{\rm lens}$ and associated $1$-$\sigma$ uncertainties measured in each individual $z$-bin with $M_c =-21.5$, for hemispherical asymmetry test.}
    \label{table:alens-NS}
\begin{tabular}{ccccc}
\hline
{\diagbox{sky}{$A_{\rm lens}$}{$z_m$}}  &0.1&  0.3 & 0.5 & 0.7 \\
\hline
\multirow{1}{*}{{North+South}} &1$\pm$ 0.15      & 1.07$\pm$ 0.07     & 1.07$\pm$ 0.05    & 0.92$\pm$0.04    \\
\multirow{1}{*}{{North}} &0.93$\pm$ 0.19      & 1.06$\pm$ 0.12     & 1.06$\pm$ 0.09    & 0.93$\pm$0.07    \\
\multirow{1}{*}{{South}} &1.15$\pm$ 0.3      & 1.1$\pm$ 0.19& 1.1$\pm$ 0.14    & 0.91$\pm$0.11   \\
\hline
\end{tabular}
\end{table*}

\cite{2004ApJ...605...14E} first report the anomalous CMB North-South asymmetry that leads to statistically significant anomalies of the CMB temperature power spectrum in the low-$\ell$ regime ($\ell\lesssim 50$). \cite{2009ApJ...704.1448H} find this asymmetry extends to much smaller scales at $\ell\lesssim 600$. Recently, such hemispherical asymmetry has been confirmed by the Planck data \citep{2014A&A...571A..23P}. In this context, the Planck lensing convergence constructed from the temperature maps may also potentially have a North-South asymmetry. We now explore the hemispherical asymmetry in the CMB lensing-LDP cross-correlation signals. By selecting LDPs from the northern and the southern ecliptic hemispheres, with the same magnitude cut of $M_c=-21.5$ as an example (without loss of generality), we find that there are essentially no systematic deviations in the scaling factor $A_{\rm lens}$ in the 4 redshift bins, $z_m=0.1, 0.3, 0.5, 0.7$, as summarized in Tab.~\ref{table:alens-NS}.

\subsection{High-redshift $A_{\rm lens}$}
\begin{table}
    \footnotesize
    \centering
    \caption{Similar to Tab.~\ref{table:alens}, but adding the data of the redshift bin of $[0.6,0.8]$ to constrain $A_{\rm lens}$, where $dof=43$, increased by 10 (as 10 $\ell$-bins).}
    \label{table:alens-z0.7}
\begin{tabular}{cccc}
\hline
Sample: &Max($M_c$) &-21.5 &Min($M_c$)\\
\hline
 $A_{\rm lens}$ & 1.1$\pm$0.03 & 1$\pm$0.03  &0.94$\pm$0.04 \\
 $\chi^2_{\rm min}/dof$ & 1.17 & 0.9 & 0.94  \\

\hline
\end{tabular}
\end{table}

As seen in Fig.~\ref{fig:simu-obs-ldp}, the measured $A_{\rm lens}$ from the individual high-redshift bin of $0.6\leq z \leq 0.8$ is lower than those from other low-redshift $z$-bins, regardless of the value of $M_c$. Thus we now compare the scaling factor $A_{\rm lens}$ derived with and without the high redshift bin as a check. As shown in Tab.~\ref{table:alens-z0.7}, adding such high-redshift data will give a slightly lower estimate on $A_{\rm lens}$ than those from the low-redshift bins (see Tab.~\ref{table:alens}), differing by only a few percent to 10\% level. In general, there is no obvious discrepancy between the highest $z$-bin and the lower ones.

In fact, due to relatively lower quality of optical data and fewer spectroscospic training sets when $z\gtrsim0.6$, the photo-$z$s of galaxies may be somewhat biasedly estimated. In the case of $M_c=-22$ which leads to more brighter galaxies are taken, the best-fit value of $A_{\rm lens}$ from the single bin of $z_m=0.7$ is $0.75\pm 0.06$ (see Fig.~\ref{fig:simu-obs-ldp}), and such underestimate is probably due to the following reasons: 1) the $r$-band magnitude may no longer be a good indicator in mass and the luminosity-halo mass relation may have a large dispersion that we have not considered; 2) the photo-$z$s and their errors of the galaxies at such high redshift might be mis-estimated and the probability distribution $p(z_{phot}|z_{obs})$ in the real situation is likely to be non-Gaussian. As illustrated in (Sun et al. in preparation), a more reliable cross-correlation signal between the CMB lensing and DESI clusters are measured if one select the clusters with more subhalos, as the redshifts of clusters could be determined more accurately. 

We also try to get support for this conjecture by matching these bright sources to a different photo-z catalog \citep{2021MNRAS.501.3309Z}  for comparison, but found no systematical difference. But considering that these two catalogs have used almost the same spectroscospic data sets  (e.g. BOSS catalog) for reference during the redshift measurements, similar systematic bias could be introduced.

In short, we need to improve our understanding of photo-$z$ measurement uncertainties and correct for photo-$z$ systematic errors in order to significantly reduce the bias. Thanks to the future release\footnote{\url{https://desi.lbl.gov/trac/wiki/ClusteringWG/LSScat/SV3}} of DESI spectroscopic data, we will be able to make more accurate cross-correlation analyses in the future. 


\subsection{Varying $\ell_{\rm min}$ and $\ell_{\rm max}$}
\begin{figure*}
    \centering
    \subfigure{
     \includegraphics[width=1\linewidth, clip]{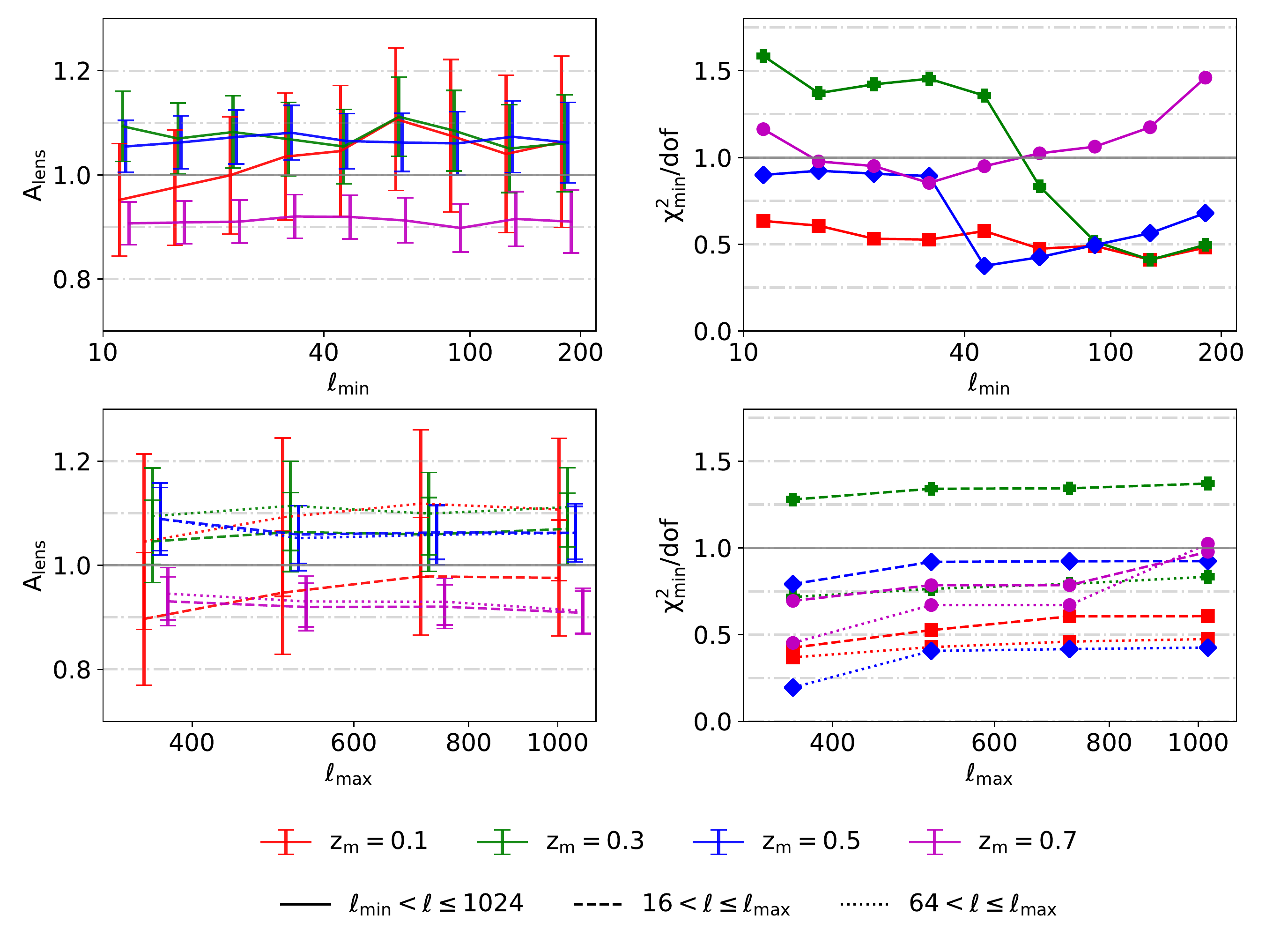}}
   \caption{Consistency check for $A_{\rm lens}$ by varying the multipole range. All the results are based on a fixed magnitude cut of $M_c=-21.5$. The solid lines in the upper left panel are the results when discarding the data points with multipoles $\ell_{min}<\ell\leq1024$. The dashed lines in the lower left panel are the results when discarding the data points with multipoles larger than $\ell_{max}$ while fixing the first data point to $\ell=16$. The dotted line has similar meaning to the dashed line, but with the first data point started from $\ell=65$. In the right two panels, we show the associated minimum reduced chi-square.} 
    \label{fig:fit_lcut}
\end{figure*}

\begin{figure*}
    \centering
    \subfigure{
     \includegraphics[width=1\linewidth, clip]{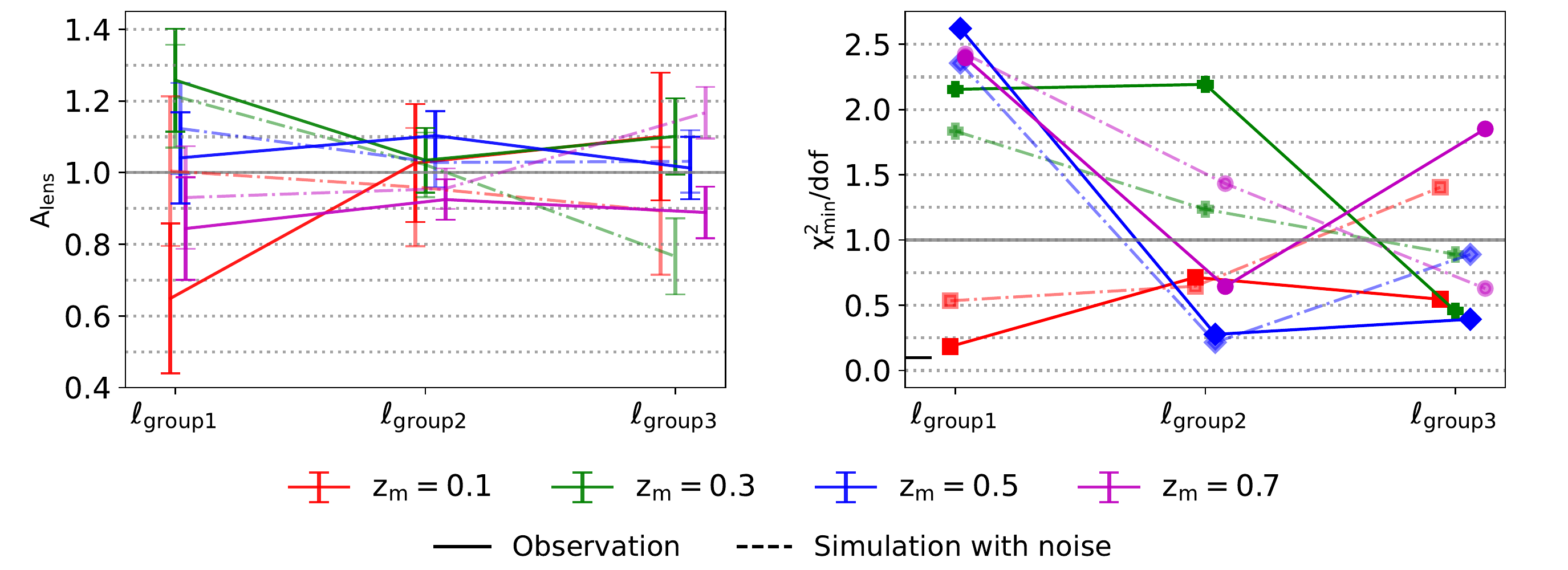}}
   \caption{Consistency check for $A_{\rm lens}$ measured within three non-overlapping multipole bins: $\ell_{group1}=$[12, 46], $\ell_{group2}=[46,256]$ and $\ell_{group3}=[256,1024]$. These two panels have similar meaning to Fig.\ref{fig:fit_lcut}. All the results are based on a fixed magnitude cut of $M_c=-21.5$. The solid lines are the results by comparing observational signals to simulation. The dashed line has similar meaning to the solid line, but using signals detected with noisy $\kappa$ map in simulation instead of observational signals.}
    \label{fig:fit_lgroup}
\end{figure*}

By varying the multipole range ($\ell_{\rm min}\leq \ell \leq \ell_{\rm max}$) in the analysis of the cross-power spectrum, we can check consistency of the scaling factor $A_{\rm lens}$ as more or fewer angular modes are included. We run two cases: i) gradually increasing $\ell_{\rm min}$ from 8 to 90, but with $\ell_{\rm max}=1024$ fixed; ii) gradually decreasing $\ell_{\rm max}$ from 1024 to 362, but fixing $\ell_{\rm min}=$ 16 or 64. As observed in Fig.~\ref{fig:fit_lcut}, the changes in $A_{\rm lens}$ are relatively small with respect to the associated statistical errors when varying the multipole range. We also note that, for the case of $z_m=0.1$, the values of $A_{\rm lens}$ when varying $l_{\rm min}$ and $\ell_{max}$ are changed across 1, from 0.9 -- 1.05, and the corresponding statistical errors are somewhat larger than other $z$-bins. This may be because of a large Poisson noise from the limited number of the galaxies in such bin ($\sim 0.28\times 10^6$ galaxies for $z_m=0.1$, see Tab.~\ref{table:DR8}).
In addition, the reduced $\chi^2$ fluctuates over all the scales when varying the multipole range, which may indicate a different scale-dependence of $C_{\ell}^{l\kappa}$ between the observed one and the template model.

Furthermore, we divide data into three non-overlapping multipole bins to check $A_{lens}$: [12, 46], [46, 256] and [256, 1024]. The results are shown as the solid lines in Fig.\ref{fig:fit_lgroup}, from which we find $A_{lens}$ is consistent. Meanwhile, we see fluctuations in both $A_{lens}$ and reduced chi-square, which might be caused by the noise in the observational lensing map. With this concern, we perform a further test with N-body simulation by adding one of the Planck noise simulation map to the ideal $\kappa$ map to mimic this effect. And similar results have been got, which are shown as the dashed lines in Fig.\ref{fig:fit_lgroup}, indicating that the noise in the lensing map is possible to disturb the cross powers at all scales. Based on these tests, we conclude that no obvious systematic bias is found for $A_{lens}$ when varying the multipole range of the observational signals.

\subsection{Adding different Galactic Masks}
\label{sec:angmask}
Based on the two facts that, the reconstructed Planck CMB lensing maps may suffer from foreground contamination especially in regions close to the Galactic plane, and the galaxy data may also be contaminated by Galactic foreground stars near the Galactic plane, we therefore add an additional mask to the maps, while combing it with the fiducial one. We find that, when adding the mask of $|{\rm Dec}|>30^\circ$ that vanishes the Galactic plane regions, the change in the factor $A_{\rm lens}$ is relatively small ($A_{\rm lens} = 1.1\pm0.08$), compared to the case of the fiducial mask together with the parameters of $M_c=21$ and $z_m=0.3$.  Also,  when adding a relatively larger mask of $|{\rm Dec}|>35^\circ$, the factor $A_{\rm lens}$ slightly increases to $1.15\pm0.1$. In the both cases, the derived $A_{\rm lens}$ are consistent with the result from the fiducial mask (${A_{\rm lens}=1.12\pm0.08}$, shown in Fig.~\ref{fig:simu-obs-ldp}), within 1-$\sigma$ level. We also find, using the samples selected by $M_c=-21.5$ with the mask of $|{\rm Dec}|>30^\circ$,  $\rm{A_{lens}}$ are found to be $1.13\pm0.14, 1.08\pm0.09, 1.11\pm0.06$ and $0.96\pm 0.05$ for $z_m=0.1,0.3,0.5$ and 0.7, respectively, consistent with the results shown in Fig.~\ref{fig:simu-obs-ldp}. We can thus conclude that, altering the region near the Galactic plane used for calculating the cross-power spectrum does not significantly change the estimate of  $A_{\rm lens}$.

\subsection{Introducing photo-$z$ bias}
\begin{figure*}[!htb]
    \centering
    \subfigure{
     \includegraphics[width=1\linewidth, clip]{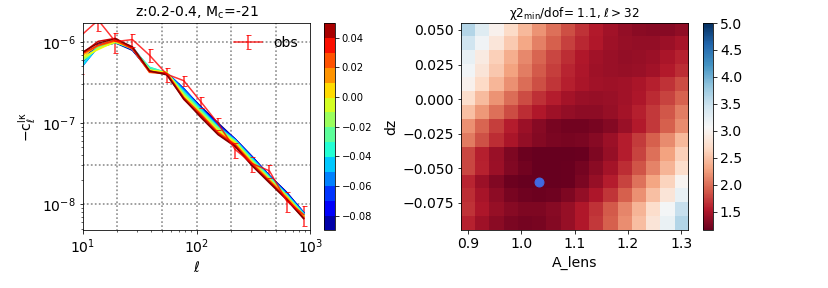}}
   \caption{Impacts of Photo-$z$ bias ($dz$) on cross-power spectrum $C_{\ell}^{\kappa l}$ and $A_{\rm lens}$ using LDPs selected in redshift bin of [0.2,0.4] by $M_c=-21$. Left panel: the color gradient lines show the resulting spectras by varying  $dz$ from -0.09 to 0.05 by 0.01 ( i.e., $[-0.09,-0.08, \cdots, 0.04, 0.05]$ in simulation, and the red solid line with error bars shows observational signal. Right panel: the reduced $\chi^2$ in the two-parameter space of $dz$ and $A_{\rm lens}$, which reaches the minimum by $dz_{\rm fit}=-0.06$ and $A_{\rm lens}=1.03$ (denoted by the blue point).}
    \label{fig:simu-zshift}
\end{figure*}

All the above analysis is based on the assumption of a Gaussian distribution for $z_{\rm p}$ centered at the true redshift $z_{\rm true}$, without any photo-$z$ bias. However, systematic photo-$z$ bias would propagate into errors in cross-correlation measurements and hence lead to a misestimate on $A_{\rm lens}$.

For checking this possibility, we focus on using LDP samples selected with $M_c=-21$ in the $z$-bin of [0.2,0.4] as an illustration. Based on the simulation data, by shifting each mock galaxy by a redshift of $dz$ that leads to a systematic redshift bias generated at $z_{\rm true}$.  We then add photo-$z$ uncertainties to those shifted galaxies and recalculate those absolute magnitudes according to $dz$, together with re-running the abundance matching process. To clearly illustrate such effects, $dz$ is varied from -0.09 to 0.05 by 0.01 in the analysis. With such new sets of mock galaxies, the measured cross-correlations are shown in the left panel of Fig.~\ref{fig:simu-zshift}. One can find that, the above photo-$z$ bias does introduce visible errors, (leading to changes of 10\% -- 20\% level relative to the case without photo-$z$ errors) in the cross-power spectrum $C_\ell^{l\kappa}$ and, consequently, the best-fit parameters are obtained, $dz_{\rm fit}=-0.06$ and $A_{\rm lens}=1.03$ for $\ell\ge33$, and $dz_{\rm fit}=-0.06$ and $A_{\rm lens}=1.06$ for $\ell\ge12$, by minimizing $\chi^2$. In other words, by introducing a negative $dz$, one may reduce some overestimate on $A_{\rm lens}$. An incorrect photo-$z$ model thus would cause a shift in measured value of $A_{\rm lens}$. 

Similarly, we obtain ($dz_{\rm fit}$, $A_{\rm lens})=(-0.05, 0.99)$, $(-0.04, 0.89)$ and $(0.04, 0.78)$ for the LDP samples selected by $M_c=-21, -21.5$ and -22 in the redshift bin of $[0.6,0.8]$, respectively. However, it is almost impossible for us to obtain an effective constraint when $M_c = -22$, since the number of such galaxy sample is too small ($n_{\rm gal}\sim 2\times 10^5$). In addition, for $z\in [0.4, 0.6]$, the best-fit $dz$, $dz_{\rm fit}$, is found to be almost zero and thus no obvious redshift bias is found. Furthermore, notice that, we can not predict the photo-$z$ biased power spectrum from mock data at the lowest $z$-bin [0.01, 0.2], since the redshifts of a large number of galaxies will become negative when shifting the galaxies by offsets $dz$.

Note that, the effects of photo-$z$ bias may also degenerate with changes in other cosmological parameters \footnote{The lensing signal detected with LSS tracers can be regarded as the matter-lensing cross power $C_l^{mk}$ multiplied by $A_{lens}b_g$, where $b_g$ is the tracer bias.  Unless we have numerical simulations for different cosmologies, we are unable to quantitatively tell the effect of cosmology on $A_{lens}$ measurements as it is coupled with $b_g$.  However, we could make a qualitative analysis of $A_{lens}$ here by simply using $c_l^{mk}$. We find that Planck cosmology leads to a slightly higher $C_l^{mk}$ than WMAP for $\ell>100$, and for $\ell<40$ it is on the opposite. And the best-fit value of $A_{lens}b_g$ is decreased by $4\%-9\%$ on the whole within our fitting range. If assuming the same $b_g$ in both cosmologies, we can imagine that  $A_{lens}$ detected under Planck cosmology would become slightly lower.  Meanwhile, we find that the chi-square between observation and the theoretical curve (with the best-fit $A_{lens}b_g$) become slightly larger in most cases under Planck cosmology.} such as $\Omega_m,\,\sigma_8$, and the tracer bias $b_g$ (as $C_\ell^{\kappa m} = A_{\rm lens} b_g C^m_\ell$), so that the ongoing DESI spectroscopic surveys are obviously necessary for precise constraints on cosmology.


\subsection{Absolute Magnitude within a certain range}
In the fiducial case, the LDP samples are generated with galaxies that are selected by their absolute magnitudes, satisfying $M_r-5 \lg h\leq M_c$. For a given redshift bin but with different $M_c$, obviously, $M_c$-selected galaxy samples are not completely independent and some of them will be overlapped. With this concern, we have performed a further test by producing LDPs for intermediate surface brightness galaxies ($-21.5\le M_r\le -21$) at $z_m =$ 0.1, 0.3, 0.5, and 0.7, respectively. The $A_{\rm lens}$ for $\ell \in [22, 724]$ are constrained to be $1.09\pm0.13, 1.15\pm0.08,1.15\pm0.06,1.04\pm0.05$, in full agreement with the fiducial case.


\subsection{Cross-Correlation between $C_\ell^{\kappa l}$ and $C_\ell^{\kappa g}$}
\begin{figure*}[!htb]
    \centering
    \subfigure{
     \includegraphics[width=0.24\linewidth, clip]{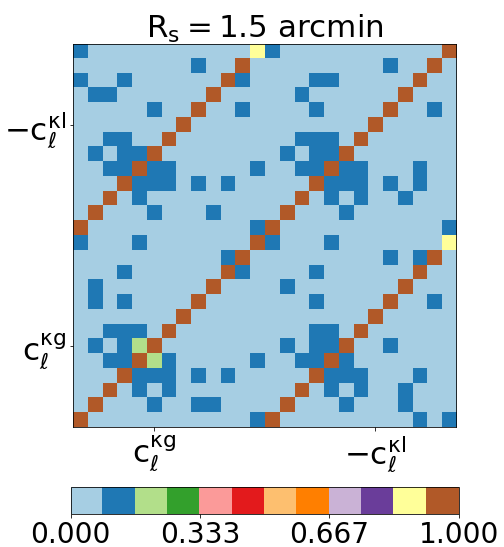}}
         \subfigure{
     \includegraphics[width=0.24\linewidth, clip]{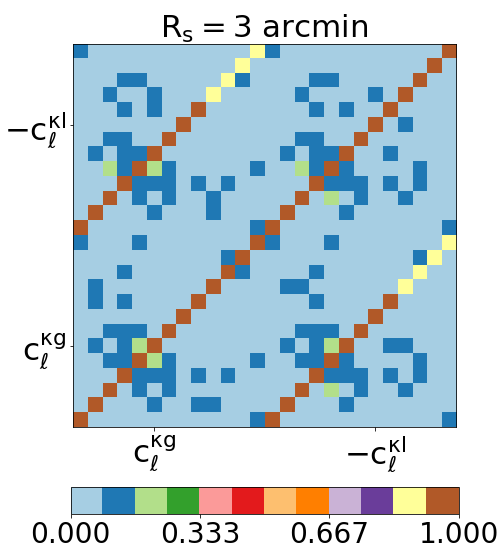}}
              \subfigure{
     \includegraphics[width=0.24\linewidth, clip]{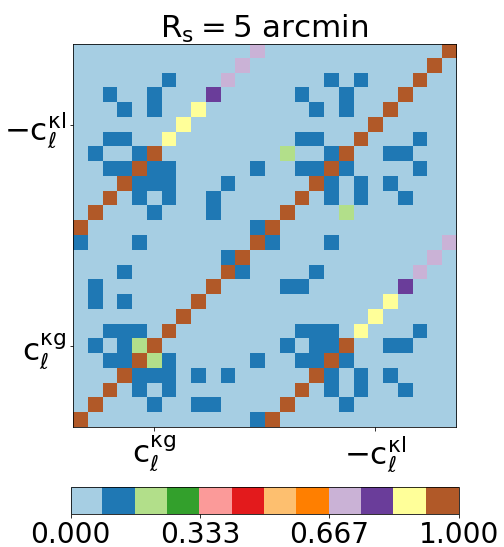}}
              \subfigure{
     \includegraphics[width=0.24\linewidth, clip]{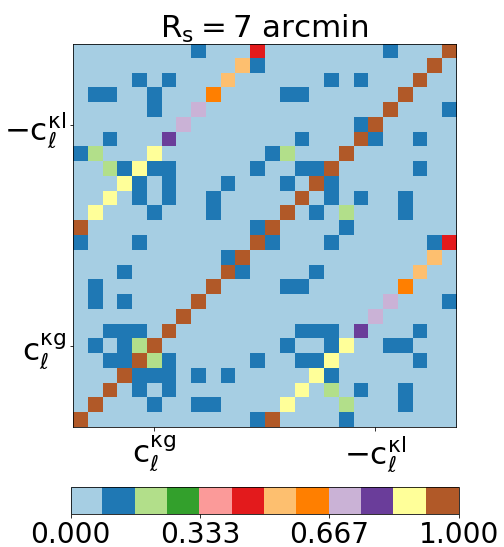}}
   \caption{Normalized covariance matrices of the vector ($C_{\ell_1}^{\kappa l},\dots, C_{\ell_n}^{\kappa l}, C_{\ell_1}^{\kappa g}, \dots, C_{\ell_n}^{\kappa g}$), calculated with varying $R_s$ from $1.5'$ to $7'$ (from the left panel to the right) in the range of $11< \ell \leq 1024$.}
    \label{fig:obs-ldp-Rs}
\end{figure*}

The cross-correlation between the lensing-LDP spectrum $C_\ell^{\kappa l}$ and the lensing-galaxy one $C_\ell^{\kappa g}$ provides a consistency check on the validity of the methodology for the CMB lensing-LDP measurements. As expected, the signal $C_\ell^{\kappa l}$ would anti-correlate with $C_\ell^{\kappa g}$ since the LDP field and galaxies as desired trace underdense and overdense regions of $\delta_m$, respectively, implying negative correlations between them. 

To do so, we calculate the auto- and cross-correlations between $C_\ell^{\kappa l}$ and $C_\ell^{\kappa g}$ for different $\ell$-bins and the corresponding normalized covariance matrices, shown in Fig.~\ref{fig:obs-ldp-Rs}. In this analysis, we consider different cases by varying $R_s$ from 1.5' to 7' but keeping the magnitude cut $M_c=-21.5$ and the redshift range [0.2.0.4] fixed for the LDP sample selection.


One can find strong anti-correlations between $C_\ell^{\kappa l}$ and $C_\ell^{\kappa g}$ whereas the absolute values of the diagonal components, $\left<C_{\ell i}^{\kappa l} C_{\ell i}^{\kappa g}\right> - \left<C_{\ell i}^{\kappa l}\right>\left<C_{\ell i}^{\kappa g}\right>$ with $i=1,\dots,n$, decrease with increasing $R_s$, as well as with increasing $\ell$. Such decreasing trend can be interpreted as the following: 1) the correlation between the overdense and underdense regions would become weaker or even disappear when significantly increasing $R_s$; 2) the stronger correlations at large scales than at small scales is due to that they may statistically have a similar spatial distribution at large scales but behave very differently at small scales.

Also, there is a transition scale ($\ell_{*}$) at which the $C_\ell^{\kappa l}$-$C_\ell^{\kappa g}$ correlation drops to below $\sim$ 0.9 and the scale depends on the value of $R_s$. We find an non-trivial but monotonic dependence of the transition scale on $R_s$, approximately $R_s$ =1.5' for the transition scale of $\theta_{*}\sim15'$ (by $\theta_{*}\approx 180^\circ/\ell_*$), $R_s =3'$ for $\sim42'$, $5'$ and $7'$ for $\sim 2^{\circ}$ and $\sim 16^{\circ}$, respectively.

In addition, the S/N for the null hypothesis over the range of $22\leq\ell\leq1024$, are $({\rm S/N})^{l\kappa}_{\rm null}=$17, 16.9, 15.7 and 14.5 for $R_s=1.5'$, 3', 5' and 7', respectively.  If calculating the S/N from the $\kappa$-galaxy correlation alone, we find $({\rm S/N})^{g\kappa,{\rm tot}}_{\rm null}=17.3$, slightly higher than the LDP case. These results thus imply LDPs can be served as a promising tracer of LSS. Moreover, one may concern whether the joint analysis of $\kappa$-galaxy and $\kappa$-LDP could enhance the S/N. To do so, we then consider the the full covariance by inclusion of auto-correlations, $\boldsymbol{C}^{l\kappa}$-$\boldsymbol{C}^{l\kappa}$ and $\boldsymbol{C}^{g\kappa}$-$\boldsymbol{C}^{g\kappa}$, a cross-correlation,  $\boldsymbol{C}^{l\kappa}$-$\boldsymbol{C}^{g\kappa}$, as shown in Fig.~\ref{fig:obs-ldp-Rs}, leading to a combined significance of $\rm( S/N)^{joint}_{null}=$ 18, 18, 18.1 and 18.2 for $R_s=1.5'$, 3', 5' and 7', respectively. If using the samples at the higher $z$-bin of $z_m=0.5$ and with $M_c=-21.5$, we find $({\rm S/N})^{l\kappa}_{\rm null}=$21.5, 21.4, 19.6 and 16.1 for $R_s=1.5'$, 3', 5' and 7', and $({\rm S/N})^{g\kappa}_{\rm null}=21.2$, resulting a combined significance of $\rm( S/N)^{joint}_{null}=$ 22, 22.2, 22.5 and 22.1 for $R_s=1.5'$, 3', 5' and 7', respectively. Thus, the use of LDP fields certainly provides an additional information on lensing detection, despite the fact that this improvement is not significant (enhanced by $\sim4\%-6\%$) due to the galaxy-LDP strong correlation. As a further check, the samples chosen in the redshift bin of $z_m=0.7$ will lead to a similar result as above. For redshifts $z_m=0.1, 0.9$ and 1.1, the $(S/N)^{\rm joint}_{\rm null}$ are found to be improved by $\sim 20\%$, 4-8$\%$ and $10\%$, respectively, compared to the lensing-galaxy case.

\section{Conclusion}
\label{sec:summary}

In this study, we have investigated the potential of utilizing LDPs (low-density points, initially proposed in \cite{2019ApJ...874....7D}) to detect the CMB lensing signal, with the aim to further verify the LDPs as a competitive tracer of LSS. By cross-correlating the Planck CMB lensing convergence map with the LDPs of different photometric redshift bins produced with galaxies selected from the DESI DR8 photo-z catalog, we have achieved an unprecedented detection significance of CMB lensing-LDP cross-spectrum of about $53\sigma$, by combining the measurements over all the multipoles $12\leq\ell\leq 1024$ and photo-$z$ bins ($0< z\le1.2$). 

For comparison, we have tested our measurements using simulated data sets, constructed specifically to precisely mimic various effects (such as sample selection, observational effects, photo-$z$ errors) of our observables: LDP catalog for DESI photometric survey and CMB convergence maps for the Planck observation. When fitting a scaling factor in the amplitude of CMB lensing-LDP cross-power spectrum, $A_{\rm lens}$, we find that, the cross-correlation amplitudes for different redshift bins and magnitude cuts are well consistent with the expected levels of cross-correlations predicted from our N-body simulations with the $\Lambda$CDM model, i.e., $A_{\rm lens}=1\pm0.12$ (1-$\sigma$ uncertainty) at $z<0.2$, $A_{\rm lens}=1.07\pm0.07$ at $0.2<z<0.4$ and $A_{\rm lens}=1.07\pm0.05$ at $0.4<z<0.6$, with the magnitude cut of -21.5 for the galaxy selection.

We also have carefully tested the various effects that could bias our result, with consistency checks on i) the hemispherical asymmetry by selecting LDPs from the northern and the southern ecliptic hemispheres, ii) selecting LDPs from the highest redshift bin, iii) scale cut by varying the the multipole range [$\ell_{min}, \ell_{\rm max}$] used in the analysis, iv) masking the regions near Galactic plane, (v) photo-$z$ bias in measurements, vi) absolute magnitude within a certain range, vii) cross-correlation between lensing-LDP and lensing-galaxy power spectra. Results are found to be consistent with the measured result from the baseline samples. No changes in our data analysis procedure have significant impacts on measured cross-correlation amplitudes, within the statistical errors. According to our study, the CMB lensing-LDP tomographic cross-correlation will be particularly important as a cosmological tool and is expected to serve as an ingredient in the framework for the analysis of cosmological parameters.

Still, many other possibilities need to be further explored. For example, the CMB lensing kernel is broad, peaked at redshift around 2. Thanks to the Wide-field Infrared Survey Explorer that allows to identify a large number of galaxies up to $z\sim1.5$ (see Tab.~1 in \cite{2019ApJS..240...30S}), therefore, we will consider a combination of optical and (near) infrared measurements to push high S/N CMB Lensing-LDP cross-spectrum measurements to a high-redshift regime. Moreover, the high-redshift cross-correlations may help to effectively break down the degeneracy among cosmological parameters and systematic nuisance parameters such as galaxy bias in LSS measurements, thus providing an accurate probe of the late-time cosmic acceleration.

\section*{Acknowledgments}
We thank Yipeng Jing for providing us the N-body simulation. We also thank valuable suggestions from the referee. The author thanks the HEALPix/healpy software package \citep{Zonca2019,2005ApJ...622..759G}. 
The author acknowledges the Korea Institute for Advanced Study for providing computing resources (KIAS Center for Advanced Computation) for this work.
F.Y.D. is supported by a KIAS  Individual Grant PG079001 at Korea Institute for Advanced Study. C.B.P. is supported by a KIAS Individual Grant PG016904 at Korea Institute for Advanced Study. This work is supported by the National Key R$\&$D Program of China (2018YFA0404504, 2018YFA0404601, 2020YFC2201600), National Science Foundation of China (11621303, 11653003, 11773021, 11890691), the 111 project, and the CAS Interdisciplinary Innovation Team (JCTD-2019-05).

\bibliography{cmbl}
\nocite{*}
\appendix \label{app}

\section{Downgrading the Planck CMB lensing maps}
 \begin{figure*}
    \centering
    \subfigure{
     \includegraphics[width=0.4\linewidth, clip]{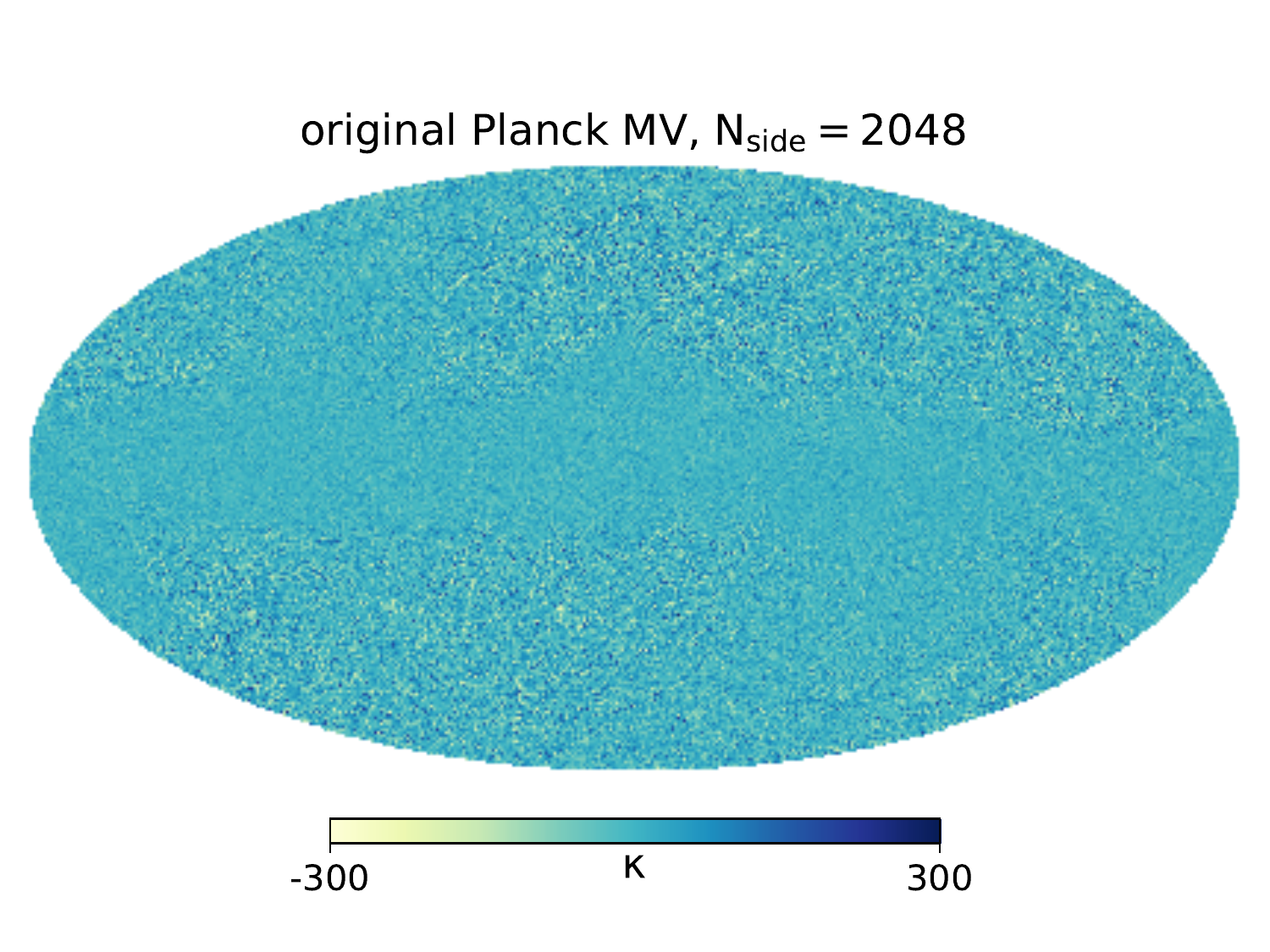}}
     \subfigure{
     \includegraphics[width=0.4\linewidth, clip]{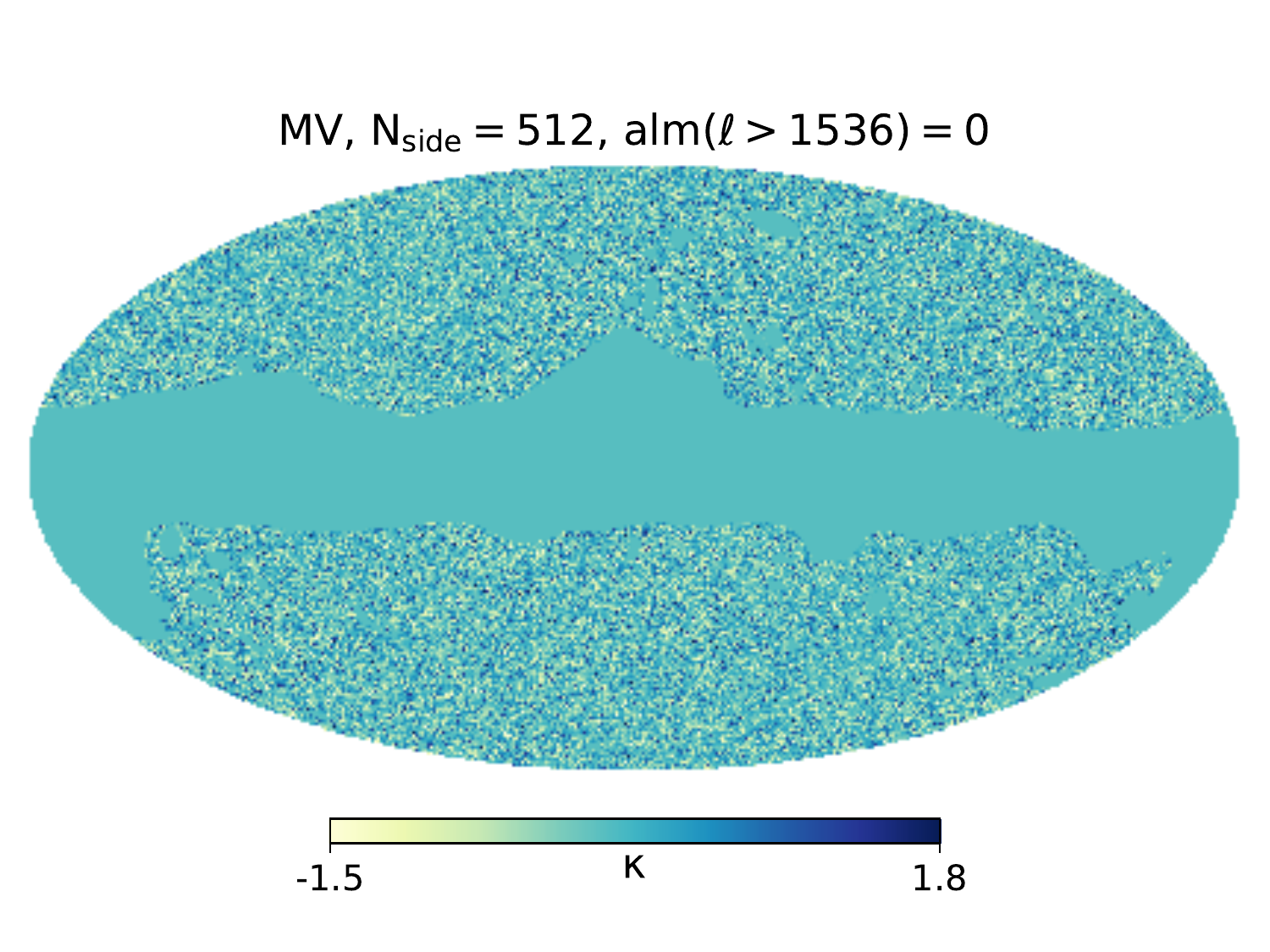}}
		\subfigure{
     \includegraphics[width=0.4\linewidth, clip]{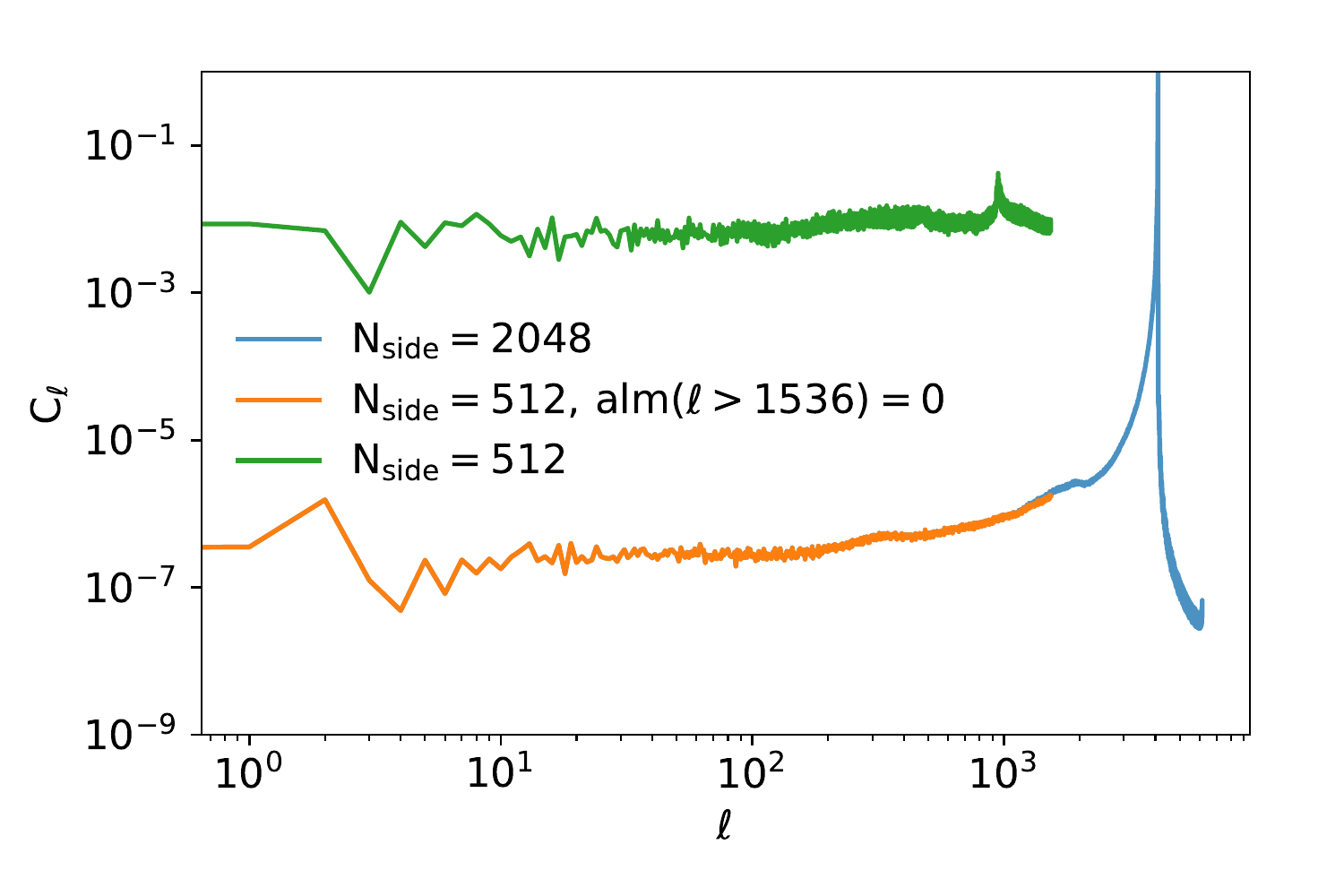}}
\caption{ Original and downgraded convergence maps using our downgrading procedure, and the corresponding power spectra. Top-left panel: the "MV" convergence map at a HEALPix resolution of $N_{\rm side}=2048$, generated from original spherical harmonic coefficients $a_{\ell m}$ provided by the Planck. Top-right panel: the masked "MV" convergence map with $N_{\rm side}=512$, produced from the low-pass filtered spherical harmonic coefficients with the cutoff at $\ell=3\times 512$.  Bottom panel: the angular power spectrum of the downgraded map of $N_{\rm side}=512$ by using our downgrading procedure (orange line), are well consistent with that of the original high-resolution map of $N_{\rm side}=2048$ (blue), without any aliasing problem. If directly downgrading the original map without using any low-pass filters, the resulting power spectrum (green) is overestimated by several orders of magnitude, which is due to the aliasing effect that leads to mapping of the noise-dominated power (a nearly divergent spike at high $\ell$) to the signal.}
    \label{fig:kappa-mapo}
\end{figure*}

In the analysis presented in this paper, we adopt the downgrading procedure as follows, in order to prevent the so-called aliasing effect that causes a mixture of signals at different scales. In principle, applying a low-pass filter to the input signal before sampling can avoid aliasing.

In general, when downgrading a map from its original high resolution to a lower one, 1) we first decompose the full-sky map into spherical harmonics ($a_{\ell m}$) of the input HEALPix resolution; 2) these coefficients are then filtered to the low one by multiplying an appropriate low-pass filter ($w_{\ell m}$) that sets all high-$\ell$ coefficients to zero; 3) then the modified filtered coefficients ($a'_{\ell m} = a_{\ell m}w_{\ell m}$) are used to directly synthesize a low-resolution map, without aliasing. 

Specifically, in order to obtain a map at the resolution of $N_{\rm nside}=512$, we can directly jump to step 2 and set $w_{\ell m} =0$ for $\ell>3\times 512$ and $w_{\ell m} =1$ otherwise as the low-pass filter, since the Planck only provides $a_{\ell m }$ up to $\ell_{\rm max} = 4096$ rather than maps. Additionally, when downgrading the binary mask ($N_{\rm side}$=2048) into a lower resolution ($N_{\rm side}$=512) in this paper, the downgraded mask is thresholded by setting pixels where the value between 0 and 1 are set to zero and the rest are set to unity so as to again generate a binary mask.  From our tests, we find such procedure can prevent the aliasing of high-$\ell$ noise power spectrum (see the spike feature at right panel of Fig.~\ref{fig:kappa-mapo}).


\section{CMB LENSING SIGNALS MEASURED WITH GALAXIES}
\label{sec:galaxy-alens}
\begin{figure*}[!htb]
    \centering
    \subfigure{
     \includegraphics[width=1\linewidth, clip]{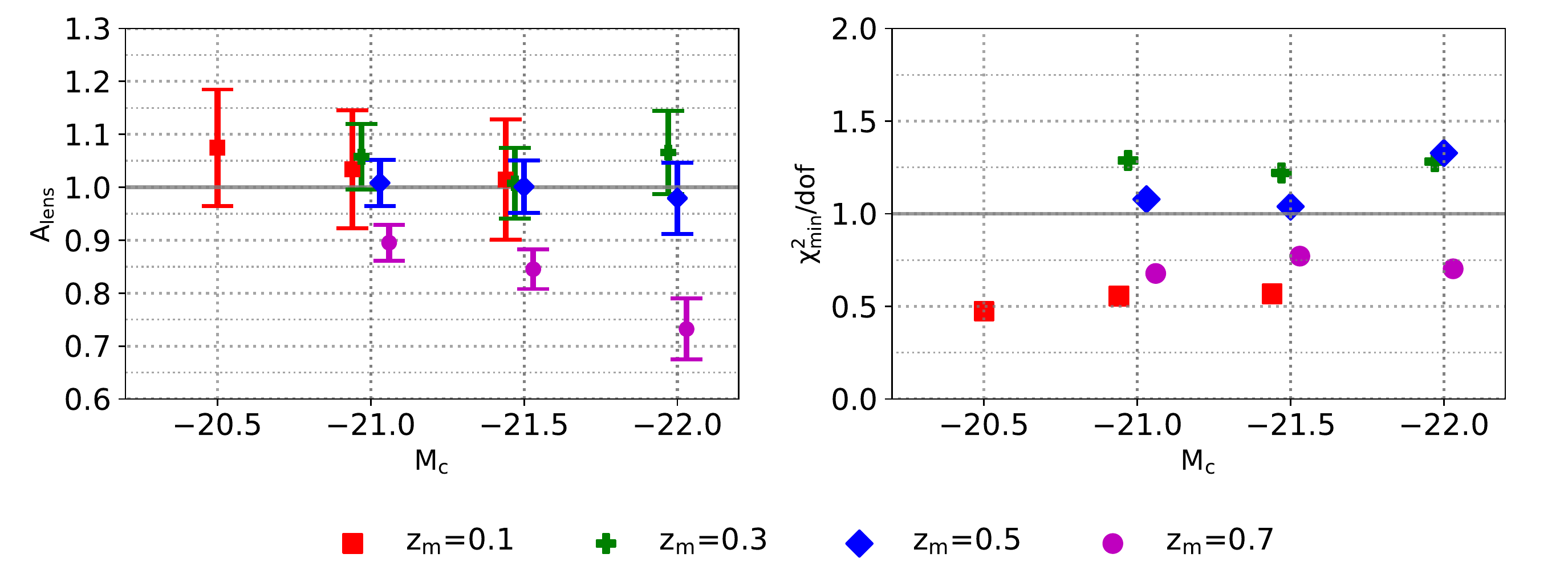}}
   \caption{Similar to Fig.\ref{fig:simu-obs-ldp}, but the measured $A_{\rm lens}$ are derived from galaxies rather than LDPs.}
    \label{fig:simu-obs-gal}
\end{figure*}

The overlap between DESI footprint and CMB surveys provides opportunities to better understand the late-time universe by using cross-correlations between galaxies and CMB lensing. Two detections have been reported so far: 1)  a cross-correlation over scales $30\le\ell\le 1000$ between Planck lensing maps and DESI-like luminous red galaxies (LRGs) selected from the DECaLS imaging is detected at a significance of 27.2$\sigma$ by~\cite{2020MNRAS.tmp.3681K}; 2)~\cite{2021MNRAS.501.1481H} place a constraint on a scaling factor for the lensing amplitude, $A_{\rm lens}$, determined by using Planck lensing convergence and the $78.6\%$ of the selected DESI objects for $z\le0.8$, and find $A_{\rm lens} =0.901\pm 0.026$, for $\Omega_m=0.275, \sigma_8=0.814$. 


 As a cross-check, we perform a cross-correlation measurement between Planck lensing and DESI galaxies. Unlike the specifically defined LRG samples used in~\cite{2020MNRAS.tmp.3681K} or the sample selected by almost all the detectable sources in \cite{2021MNRAS.501.1481H}, we select the volume-limited galaxy samples at each photo-$z$ bin by using different absolute magnitude cuts, same as described in the main text. The constraints on $A_{\rm lens}$ for the 4 $z$-bins are shown in Fig.~\ref{fig:simu-obs-gal}, which are basically compatible to LDP results (see Fig.~\ref{fig:simu-obs-ldp}). Furthermore, the galaxy samples at  $z$-bin $0.6\leq z\leq 0.8$ yield an systematic underestimate on $A_{\rm lens}$ at beyond the 2-$\sigma$ level, relative to the theoretical model from our simulation. The obvious underestimate for the brightest galaxy sample of $M_c=-22$ is consistent with that of LDPs. This may be because of low-quality measurements of photo-$z$  and the lack of galaxies at this high redshift bin. By combining the different redshift slices with or without the highest bin, the measured $A_{\rm lens}$ from different $M_c$ are summarized in Tab.~\ref{table:alens-gal}. The inclusion of galaxy samples from the highest $z$-bin indeed clearly lower the estimated value of $A_{\rm lens}$.

 

\begin{table*}[!htb]
    \footnotesize
    \centering
    \caption{Best-fit $A_{\rm lens}$ with $1$-$\sigma$ uncertainties from different magnitude cuts $M_c$, with and without inclusion of the $z$-bin of [0.6,0.8], and the corresponding $\chi^2_{\rm min}/dof$ for each case.}
    \label{table:alens-gal}
\begin{tabular}{ccccccccccc}
\hline
&& \multicolumn{4}{c}{$ A_{\rm lens}$}&\multicolumn{4}{c}{$\rm{\chi^2_{min}/dof}$}\\  
choice of $z_m$ &  &$Max(M_c)$ &-21.5 &$Min(M_c)$ &&&$Max(M_c)$ &-21.5 &$M in(M_c)$\\
\hline
\multirow{1}{*}{{w/o $z_m=0.7$}} & &1.03$\pm$0.03 & 1$\pm$0.04  &1.02$\pm$0.05    &&& 0.95 & 0.94& 1.06   \\
\multirow{1}{*}{{w $z_m=0.7$}} & &0.96$\pm$0.02 & 0.92$\pm$0.03  &0.9$\pm$0.04    &&& 0.88 & 0.9 & 0.97   \\
\hline
\multicolumn{10}{l}{Similar to Tab.~\ref{table:alens}, but the constraints are from galaxies alone.}&\\
\end{tabular}
\end{table*}
\end{document}